\documentclass{aa}  

\newif\ifboldtext
\boldtextfalse

\usepackage{placeins}

\usepackage{xcolor}
\usepackage[colorlinks=true,linkcolor=blue,citecolor=blue]{hyperref}
\usepackage{graphicx}
\usepackage{comment}

\usepackage{subcaption}
\usepackage{txfonts}

\usepackage[para,online,flushleft]{threeparttable}

\begin{document}

   \title{Dispersion-supported galaxy mass profiles with convolutional neural networks}

   \author{J. Sarrato-Alós
          \inst{1}\fnmsep\inst{2},
          C. Brook \inst{1}\fnmsep\inst{2},
          A. Di Cintio\inst{1}\fnmsep\inst{2},
          J. Expósito-Márquez
          \inst{1}\fnmsep\inst{2},
          M. Huertas-Company\inst{1}\fnmsep\inst{2}\fnmsep\inst{3}
          \and
          A. V. Macciò\inst{4}\fnmsep\inst{5}\fnmsep\inst{6}
         }

   \institute{Instituto de Astrofísica de Canarias (IAC), Calle Via Láctea s/n, E-38205 La Laguna, Tenerife, Spain\\
   \email{jorge.sarrato@iac.es}
         \and
             Universidad de La Laguna, Avda. Astrofísico Fco. Sánchez s/n, E-38206 La Laguna, Tenerife, Spain
             \and
             Université de Paris, LERMA - Observatoire de Paris, PSL, Paris, France
             \and
             New York University Abu Dhabi, PO Box 129188, Abu Dhabi, United Arab Emirates
             \and
             Center for Astrophysics and Space Science, New York University Abu Dhabi, Abu Dhabi, PO Box 129188, Abu Dhabi, UAE
             \and
             Max-Planck-Institut für Astronomie, Königstuhl 17, D-69117~Heidelberg, Germany}

   \date{Received x xx, xxxx; accepted x xx, xxxx}

  \abstract

   {Determining the dynamical mass profiles of dispersion-supported galaxies is particularly challenging due to projection effects and the unknown shape of their velocity anisotropy profile. Traditionally, this task relies on time-consuming methods that require profile parameterisation and the assumption of dynamical equilibrium and spherical symmetry.}
   {Our goal is to develop a machine-learning algorithm capable of recovering dynamical mass profiles of dispersion-supported galaxies from line-of-sight stellar data. }
   {We trained a convolutional neural network model using various sets of cosmological hydrodynamical simulations of galaxies. By extracting projected stellar data from the simulated galaxies and feeding them into the model, we obtained the posterior distribution of the dynamical mass profile at ten different radii. Additionally, we evaluated the performance of existing literature mass estimators on our dataset.}
   {Our model achieves more accurate results than any literature mass estimator while also providing enclosed mass estimates at radii where no previous estimators exist. We confirm that the posterior distributions produced by the model are well calibrated, ensuring they provide meaningful uncertainties. However, issues remain: the method's performance is less good when trained on one set of simulations and applied to another, highlighting the importance of improving the generalisation of machine-learning methods trained on specific galaxy simulations.}
   {}


   \titlerunning{Galaxy mass profiles with CNNs}
   \authorrunning{J. Sarrato-Alós et al.}

   \maketitle

\section{Introduction}

Understanding the dynamical mass distribution within galaxies is crucial for testing the cold dark matter paradigm. In rotation-supported galaxies, mass is inferred from rotational velocities, whereas dispersion-supported galaxies require kinematic studies based on spectroscopic data to trace stellar motions. These measurements are subject to projection effects, making it difficult to reconstruct the full 3D structure of the galaxy.

Dynamical modelling provides a robust approach to addressing these challenges. Advanced techniques, often combined with Monte Carlo simulations, yield robust mass estimates with well-calibrated uncertainties. However, these models are still affected by the mass-anisotropy degeneracy; unknown velocity anisotropies can bias the inferred mass distribution, with some methods more sensitive to these effects than others. Moreover, the computational expense and the need for careful system modelling further complicate the process.

In contrast with dynamical modelling,  simple formulae have been developed \citep[e.g.][]{Walker_estim, Wolf_estim, Amorisco_estim, Campbell_estim, Errani_estim} to estimate the dynamical mass of galaxies enclosed within specific radii; in these approaches, velocity anisotropy and/or other factors have been found to minimise uncertainty. These estimators rely on the line-of-sight velocity dispersion of the stellar component and on the half-light radius, aiming to obtain an unbiased mass estimate that holds for a broad range of galaxy properties and is minimally affected by line-of-sight projection effects.

In recent years, machine learning has proven to be an effective tool for analysing complex data and uncovering underlying patterns. These methods have been applied to various astrophysical problems, including dynamical mass estimation and density profile determination using line-of-sight data. For example, \cite{fabio_iocco} used convolutional neural networks (CNNs) to analyse mock photometry and interferometry images from cosmological hydrodynamical simulations of spiral galaxies, successfully inferring 20 points of their dynamical mass profiles. Similarly, \cite{julen} employed 2D probability distribution functions (PDFs) of projected stellar positions and kinematics as inputs to a CNN to estimate the inner slopes (150 pc from the centre) of dark matter density profiles in dwarf galaxies. This approach resulted in a model that was able to differentiate between cusps and cores in cosmological hydrodynamical simulations. \cite{Ho_2019_clusters} applied a CNN model with a similar input structure based on PDFs of the positions and velocities of galaxies to estimate the dynamical masses of galaxy clusters. Also, \cite{Nguyen_2023} developed a graph neural network capable of accurately recovering the dark matter density profiles of mock spherical, dynamically equilibrated dwarf galaxies.

In this study we built a CNN based on that used in \cite{julen} and trained it on a mixed dataset of cosmological hydrodynamical simulations of dispersion-supported galaxies spanning a broad range of stellar masses and other galactic properties. Our model uses projected stellar information to estimate the dynamical mass profile of galaxies, providing the posterior distribution of the values of ten points of the dynamical mass profile of the galaxy. We compared the capabilities of our model with five existing literature mass estimators \citep{Walker_estim, Wolf_estim, Amorisco_estim, Campbell_estim, Errani_estim}. We find our model provides improved scatter in the estimation of dynamical mass, on top of its ability to provide new points of the mass profile at radii where no literature mass estimators have been proposed.

The structure of this article is the following: In Sect. \ref{sec:simulations} we describe the simulations used to test the literature mass estimators and train the CNN, and also the process we followed to prepare the dataset to be inputted into the model. In Sect. \ref{sec:massestimators} we introduce the existing mass estimators in the literature, and briefly explaining their derivation and assumptions. We describe our CNN model in Sect. \ref{sec:CNN}, where we also present its architecture and explain the structure of inputs and outputs. In Sect. \ref{sec:results} we present the results of our work. In Sect. \ref{sec:res-massestimators} we show the performance of literature mass estimators on our dataset of simulated galaxies, and then in Sect. \ref{sec:res-CNN} we compare them with the results of our model. We summarise our conclusions in Sect. \ref{sec:conc}.

\section{Simulation dataset}
\label{sec:simulations}

Training a CNN to estimate galaxy dynamical mass requires an extensive dataset of realistic simulated galaxies from which we can extract stellar data to use as input for the network, as well as calculate the enclosed mass profile we aim to estimate. We decided to use cosmological hydrodynamical simulations over idealised simulations, choosing to train our model on a more realistic dataset for which we have limited sampling capability, rather than on a less realistic dataset that we can generate and sample faster and more uniformly.

Since we are restricted to a dataset of previously run simulations, we aimed for it to sample the space of galaxy properties as completely as possible, providing a large enough number of galaxies for the study. We found a set of simulations based on the Numerical Investigation of a Hundred Astrophysical Objects \citep[NIHAO;][]{nihao_original} project  that fulfils these requirements. In order to assess possible biases in the model, we also selected other suites of simulations to compare the performance of the model. For that, we also made use of publicly available data for the AURIGA \citep{AURIGA_release} and the Feedback In Realistic Environments-2 \citep[FIRE-2;][]{FIRE2_release} simulations.

\subsection{The NIHAO project}
\label{sec:NIHAO}

The NIHAO project consists in a series of cosmological hydrodynamical zoom-in simulations run with the cosmology parameters of \cite{planckcosmo}: H$_{\text{0}}$ = 100h km s$^{\text{-1}}$ Mpc$^{\text{-1}}$ with h = 0.671, $\Omega_{\text{m}}$ = 0.3175, $ \Omega_{\Lambda}$ = 0.6824, $\Omega_{\text{b}}$ = 0.049, and $\sigma_{\text{8}}$ = 0.8344. The galaxy formation model includes ultraviolet heating, ionisation and metal cooling \citep{shen}. Star formation and feedback follow the model used in the Making Galaxies In a Cosmological Context \citep[MaGICC;][]{stinson13} simulations, reproducing galaxy scaling relations over a wide mass range \citep{brook12b}. The density threshold for star formation is $n_{\rm th}$$>$$10.3 \rm cm^{-3}$ and a \citet{Chabrier03} initial mass function is used. Energy from stars is injected into the interstellar medium, modelled using blast-wave supernova feedback \citep{stinson06} and early stellar feedback from massive stars. Particle masses and force softenings allow the mass profile to be resolved to below 1$\%$ of the virial radius, which ensures that galaxy half-light radii are well resolved with spatial resolution varying from 100 pc through 800 pc for the most massive galaxies.

\subsection{The AURIGA project}
\label{sec:AURIGA}

The cosmological parameters used in the simulations from the AURIGA project come from \cite{planckcosmo}, just like in the NIHAO project.

The galaxy formation model implemented in simulations from the AURIGA project includes magnetic fields, primordial line metal cooling that takes self-shielding into  account, stellar feedback and thermal feedback from radio and quasar accretion modes of black holes. Star formation is regulated by the Kennicutt-Schmidt law \citep{KS} and a \cite{Chabrier03} initial mass function. The resolution of publicly available simulations ranges from 200 pc to 400 pc.

The AURIGA simulation suite effectively reproduces the properties of real galaxies, particularly Milky Way-like systems. \cite{2017auriga} demonstrates that the simulated galaxies include thin and thick disk components similar to those observed in the Milky Way. The simulations also produce realistic rotation curves and star formation histories. Additionally, the galaxies follow the expected stellar mass–halo mass relation and accurately match the Tully-Fisher relation.

\subsection{The FIRE project}
\label{sec:FIRE}

The FIRE project includes simulations with different cosmologies. Some of them follow \cite{planckcosmo}, and others adopt the cosmology from the Assembling Galaxies Of Resolved Anatomy \citep[AGORA;][]{AGORA_cosmo} project: H$_{\text{0}}$ = 100h km s$^{\text{-1}}$ Mpc$^{\text{-1}}$ with h = 0.702, $\Omega_{\text{m}}$ = 0.272, $ \Omega_{\Lambda}$ = 0.728, $\Omega_{\text{b}}$ = 0.0455, and $\sigma_{\text{8}}$ = 0.807.

The FIRE suite galaxy formation model implements stellar and super-massive black hole feedback, magnetic fields, cosmic rays, photo-ionisation and photo-heating. They reach very high spatial resolution with softening lengths in the order of 10 parsecs.

The FIRE simulations, particularly the FIRE-2 suite, have been shown to effectively reproduce several key properties of real galaxies. \cite{Hopkins_2018} demonstrate that FIRE-2 galaxies align with observed morphologies, interstellar medium structure, star formation histories, rotation curves, metallicities and stellar masses.

\begin{figure}[ht]
    \centering
    \includegraphics[width = \columnwidth]{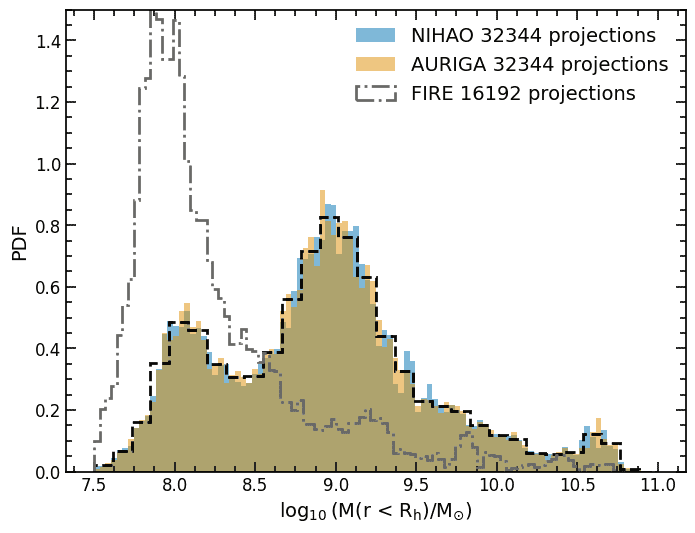}
    \caption{Distribution of dynamical mass enclosed within the projected half-light radius ($R_{\rm h}$) of dispersion-supported galaxies for each suite of simulations. Blue and orange histograms show, respectively, the distributions of NIHAO and AURIGA galaxy projections with 100 bins, while the dashed black line histogram shows their distribution at 30 bins, which we made equal in the two suites by dropping excessive projections from each dataset. The dashed grey lines show the distribution of FIRE galaxy projections inside the mass range covered by NIHAO and AURIGA.}
    \label{fig:mrdhist}
\end{figure}

\subsection{Training dataset}
\label{sec:train-set}

Our dataset comprises 496 zoom-in cosmological simulations extracted from the NIHAO project, 66 zoom-in simulations from the AURIGA project and 17 zoom-in simulations from the FIRE project. Besides the fiducial NIHAO dataset, we included simulations of NIHAO galaxies with different stellar formation density thresholds, $n_{\rm th}$ \citep{nihao_nth}, NIHAO galaxies that have black holes \citep{nihao_bh}, and high resolution runs of six NIHAO simulations \citep{NIHAO_UHD}. Due to the limited number of FIRE simulations, galaxies from this suite were only used to test the model and not for the training.

For all simulations in the NIHAO+AURIGA dataset, we selected all galaxies with stellar masses ranging from $M_{*}=10^{5.5} \mathbf{M_{\odot}}$ to $M_{*}=10^{11} \mathbf{M_{\odot}}$ that contain at least 50 star particles and whose mass fraction in high resolution particles is larger than 95\%. Out of these galaxies, we selected those that are considered dispersion-supported systems according to the criterion $\kappa_{\textrm{co}} < 0.5$. The quantity $ \kappa_{\textrm{co}}$ was introduced in \cite{kappa_salva}. It measures the kinetic energy fraction invested in ordered rotation and is defined as

\begin{equation}
    \label{eq:kappa}
    \kappa_{\textrm{co}} = \frac{1}{K_{\textrm{star}}} \sum_{i}^{r < \textrm{30 kpc} \;\; ; \;\; j_{\textrm{z,i}}>0} \frac{m_{\textrm{i}}}{2} \left( \frac{j_{\parallel,\textrm{i}}}{R_{\perp,\textrm{i}}} \right)^{2}
,\end{equation}where $K_{\textrm{star}}$ is the total kinetic energy of stars, and the sum is over all star particles within 30 kpc of the centre of the galaxy and with positive values of the specific angular momentum ($j_{\parallel}$) along the direction of the total angular momentum of the stellar component of the galaxy. $m$ is the mass of each stellar particle, and $R_{\perp}$ is the distance from the star to the rotation axis of the stellar component. We further improved the quality of the dataset by manually removing ongoing mergers and heavily disrupted galaxies identified through visual inspection. Applying these filters resulted in a sample of 1976 galaxies from the NIHAO project, 1066 from the AURIGA project, and 260 from the FIRE project.

\begin{figure}[ht]
\centering\includegraphics[width=\columnwidth]{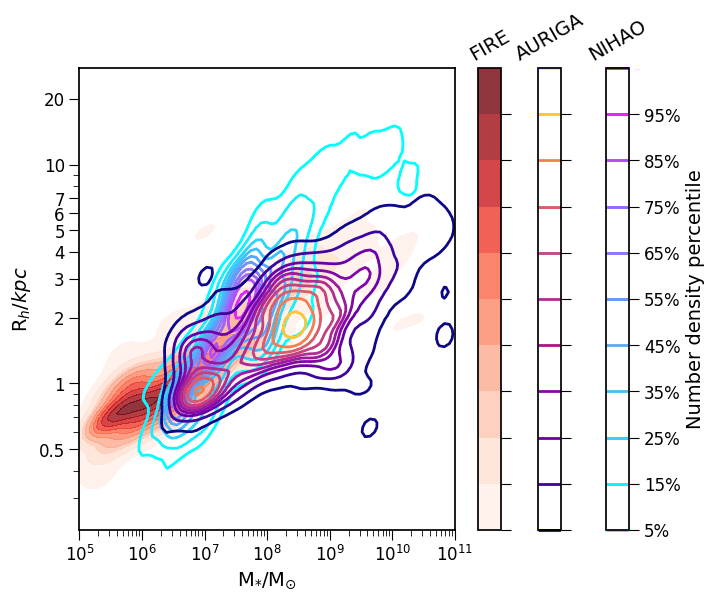}
    \caption{Number density contours of projections of galaxies of NIHAO, AURIGA, and FIRE simulation suites in the space of the stellar mass--projected half light radius.}
    \label{fig:Ms_vs_rh}
\end{figure}

We sampled stars randomly from each galaxy with a sample size between 200 and $10^{4}$, depending on the total number of star particles in the original galaxy. For galaxies with fewer than 200 star particles, we selected all of them. Then we project the stellar data of galaxies along 64 lines of sight uniformly distributed along the unit sphere. We carried on this process for galaxies of all three simulations in order to construct datasets we could use to test the performance of models over all lines of sight. For each projection we also calculated the projected half-light radius (half-stellar number\footnote{To avoid relying on any stellar flux model, we assumed that all stars have the same luminosity. Consequently, when we mention the half-light radius of the simulated galaxies, we are referring to their half-stellar number radius.}) and the line-of-sight velocity dispersion. Since the datasets from NIHAO and AURIGA simulations are fairly large ($\sim$10$^{5}$), we decided to cut them by selecting the same number of projections from each suite for each bin of dynamical mass enclosed within the half-light radius, dropping the excess projections. We used 30 bins between dynamical enclosed masses 10$^{7.5}$ and 10$^{11}$. This process enabled a well-balanced training of the neural network on a mixed dataset of AURIGA and NIHAO galaxies. We show the resulting mass distributions in Fig. \ref{fig:mrdhist}. In Fig. \ref{fig:Ms_vs_rh} we show the contours of the distribution in the space of stellar mass--projected half-light radius for all simulations.

The final training dataset consists in a total of 32344 projections of galaxies from the NIHAO project and the same number from the AURIGA project. We also obtained a dataset of 16640 projections of galaxies from the FIRE project within the same range of dynamical masses (within the projected half-light radius) as the NIHAO and AURIGA datasets. When training our neural network we made sure that all the projections of an individual galaxy lie within either the training set or the validation set, avoiding the validation of the model on data extracted from a galaxy that had also been used to train the model.

\section{Mass estimators}
\label{sec:massestimators}

\begin{figure*}[ht]
        \includegraphics[width = \textwidth]{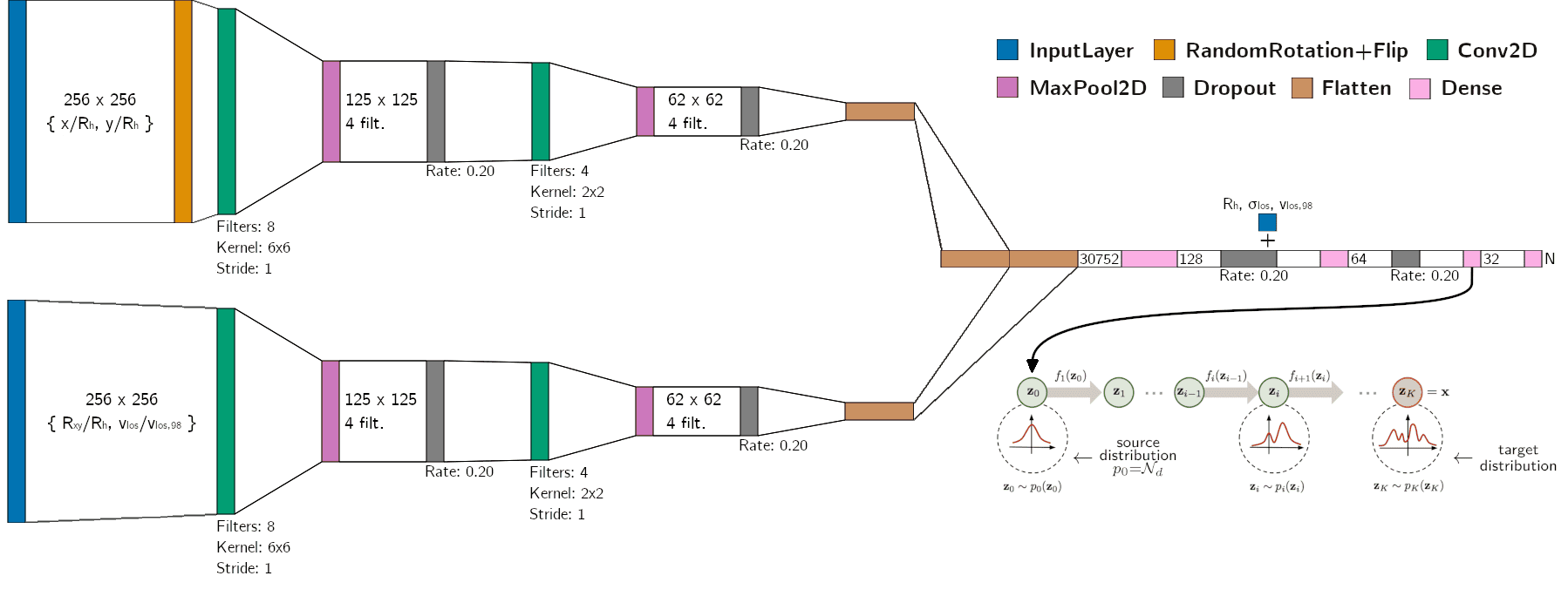}
  \caption{Scheme of our CNN architecture. The CNN extracts the spatial and dynamical information of the galaxy from the projected stellar data and compresses it through a series of convolution and pooling operations. The CNN joins the information of the spatial and dynamical branches and further reduces the dimensionality for producing an N parameter output, representing the value estimated for the dynamical mass of the galaxy enclosed within N different radii. After training the CNN, the 32 neurons of penultimate layer are used as inputs to train a normalising flow model. The flow model learns a series of transformations to a N-dimensional Gaussian PDF, which are conditioned on the inputs, and outputs a posterior N-dimensional joint PDF for the N enclosed masses.}
     \label{fig:cnn_architecture}
\end{figure*}

Several works in the literature have developed simple formulae for calculating galaxy dynamical masses enclosed within different radii from projected quantities only. These estimators make use of the projected half-light radius ($R_{\textrm{h}}$) of the galaxy and the line-of-sight velocity dispersion of the stellar component (i) in all the galaxy, (ii) at a specific distance from the centre, or (iii) within a given radius. We introduce here the most common mass estimators that we tested on our simulations in Sect. \ref{sec:res-massestimators}.

The first such mass estimator is that of \cite{Penarrubia_estim}. They find that in the case of isotropic ($\beta(r)=0$) stellar components following a King profile inside a \cite{NFW} halo, the error of using the central stellar component's projected velocity dispersion for calculating the dynamical mass of galaxies is minimised within the King core radius ($R_{c}$). Their mass estimator takes the form $M(<R_{c}) = 1.44G^{-1}R_{c}\sigma_{los}^{2}(0)$.

\cite{Walker_estim} made the less restrictive assumption that the anisotropy profile is flat throughout all the galaxy, and that the stellar component follows a Plummer spherical profile. They argued that the error on mass estimation introduced by the constant anisotropy is small when using the estimator $M(<R_{h})=2.5G^{-1}R_{h}\left\langle \sigma_{los}^{2} \right\rangle$ that uses the line-of-sight stellar velocity dispersion over the whole galaxy.

\cite{Wolf_estim} derived analytically a new mass estimator that minimises the effect of a non-uniform, three-parameter anisotropy profile when calculating the mass within $r_{3}$, the radius where the log-slope of the stellar density profile is 3. Through a series of approximations they transform their mass estimator in order to apply it to calculate the mass within $4/3R_{h}$: $M(<4/3R_{h})=4G^{-1}R_{h}\left\langle \sigma_{los}^{2} \right\rangle$. 

For stellar components modelled using the Michie-King distribution function, \cite{Amorisco_estim} find $M(<1.67R_{h})=5.85G^{-1}R_{h}\sigma_{los}^{2}(R_{h})$\footnote{Although this mass estimator uses the value of the $\sigma_{los}^{2}$ radial profile at $R_{h}$, we used the averaged value $\left\langle \sigma_{los}^{2}(<R_{h}) \right\rangle$ due to the limited number of stellar tracers of several galaxies in the dataset.}.
\cite{Campbell_estim} used the APOSTLE simulations \citep{apostle} to derive a mass estimator for larger radii $M(<1.77R_{h})=5.99G^{-1}R_{h}\left\langle \sigma_{los}^{2}(<1.04R_{h}) \right\rangle$.

\
For a similar radius, \cite{Errani_estim} find that $M(<1.8R_{h})=3.5\times1.8G^{-1}R_{h}\left\langle \sigma_{los}^{2}\right\rangle$ minimises the errors introduced by the shape of the tidally stripped dark matter halo (cusp or core) in which lies a Plummer sphere stellar component.

Some mass estimators have been tested on cosmological simulations. Apart from their own estimator, \cite{Campbell_estim} tested Walker's and Wolf's mass estimators on fully cosmological hydrodynamical simulations from the APOSTLE project. \cite{gonzalezsamaniego} tested Walker's, Wolf's, and Campbell's mass estimators on 10 dwarf galaxy cosmological hydrodynamical simulations from the FIRE-2 project \citep{fire}. The two studies obtained different results for the biases and specially for the scatters of the three mass estimators.

\section{Neural network}
\label{sec:CNN}

We used the python package pytorch \citep{paszke2019pytorch} to construct a CNN based on the one used in \cite{julen}.

\subsection{Input}
\label{sec:input}

The inputs of our neural network must represent the line-of-sight projected dynamical data of stars in galaxies in an uniform way such that we can use it to represent the data of both smaller and larger galaxies with enough detail. At the same time, the inputs must include information about the size of the galaxies, since galaxy size correlates with mass.

We constructed the inputs using the x and y line-of-sight projected positions of stars in the sky, and their velocities along the line of sight ($v_{los}$), for each projection of each galaxy in the dataset. Additionally, we used the projected half-light radius ($R_{h}$), the 98th percentile of the line-of-sight velocity ($v_{los, p98}$), and the line-of-sight velocity dispersion of stars ($\sigma_{los}$). We calculated these quantities for each projection using only the stars in the sub-sample corresponding to the projection, rather than all stars in the galaxy.

We chose to use two different 2D arrays with dimensions 256x256 as inputs. The arrays represent samples of 2D joint PDFs of projected stellar data. We constructed the PDFs using bivariate kernel density estimation (KDE), which is described in Appendix \ref{app:kde}. The input arrays were as follows:

\begin{itemize}
    \item Input 1: A 2D PDF of the projected positions of stars in the sky, normalised by $R_{h}$ (the two dimensions of the array are \{x/$R_{h}$, y/$R_{h}$\}). We sampled the PDF in a grid between -2.4 and 2.4 in both dimensions to create a 256x256 array.
    \item Input 2: A 2D PDF in the space \{$R_{xy}$/$R_{h}$, $v_{los}$/$v_{los, p98}$\} (where $R_{xy} = \sqrt{x^{2} + y^{2}}$). We sampled the PDF in a grid between -2.4 and 2.4 in the $R_{xy}$/$R_{h}$ dimension and between -1 and 1 in the $v_{los}$/$v_{los, p98}$ dimension, obtaining a 256x256 array.
    \item Input 3: Since literature mass estimators make use of line-of-sight velocity dispersion $\sigma_{los}$ and projected half-light radius ($R_{h}$), we decided to also input these values to the final layers of the network. On top of that, we included the value $v_{los, p98}$, as we used it for scaling one dimension of the array in input 2. We structured these data in a 1D array of size 3.
\end{itemize}

\subsection{Architecture}
\label{sec:architecture}

\begin{figure}[ht]
    \centering
    \includegraphics[width = \columnwidth]{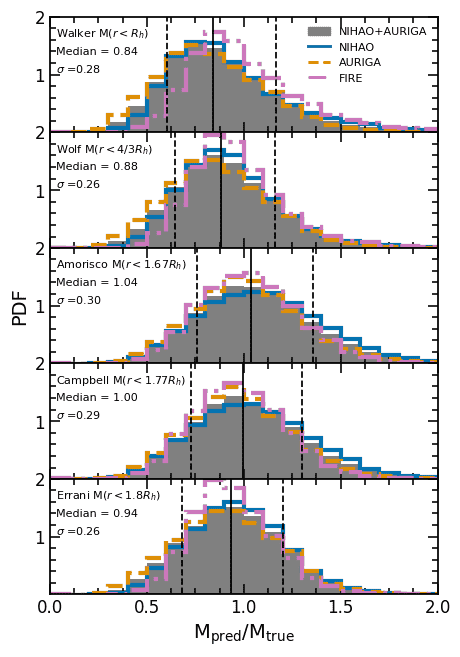}
    \caption{Probability density function of the ratio between the enclosed mass calculated by the literature mass estimators and the true enclosed mass for all dispersion-supported simulated galaxies in our training dataset of mixed NIHAO and AURIGA simulations (solid histogram), and for galaxies from each simulation project separately (empty histograms with different line styles). In the case of the combined NIHAO and AURIGA dataset, the median ratio is represented by a vertical solid line, and percentiles 16th and 84th are depicted by dashed lines for the mixed dataset. The values of the median ratio and the 1$\sigma$ dispersion range are shown. Each row represents the results from a different mass estimator.}
    \label{fig:full-estimators}
\end{figure}

The architecture of the CNN is shown in in Fig. \ref{fig:cnn_architecture}. It is formed by two branches that take as inputs both arrays described in Sect. \ref{sec:input}. The first input (projected position map \{x/$R_{h}$, y/$R_{h}$\}) is randomly rotated and flipped when fed into the model as a technique to improve rotational invariance and to avoid over-fitting. The arrays were passed twice through a combination of convolutional layer plus pooling layer, and then we applied a dropout layer to minimise potential over-fitting. After this, the results of both branches were flattened and concatenated, and we used a series of fully connected dense layers until we reduced the output to N values, where N is the number of mass profile points we intended to fit (in this case, 10). We again applied dropout between specific layers. Also, we concatenated the projected half-light radius ($R_{h}$), the line-of-sight velocity dispersion of the galaxy ($\sigma_{los}$), and the 98th percentile of line-of-sight velocities ($v_{los, p98}$) to the outputs of the first dropout layer.\\

We utilised the outputs of the CNN to train a normalising flow model \citep{normflows} with the Python package ltu-ili \citep[Learning the Universe Implicit Likelihood Inference;][]{ltuili}. Specifically, the 32 values from the penultimate layer of the CNN were employed as inputs to train a masked autoregressive flow \citep[MAF;][]{maf} for neural posterior estimation. The model learns a series of transformations conditioned on the input data, applied to a base Gaussian distribution, to produce a posterior distribution for the target variables (dynamical enclosed masses in this study). The MAF model outputs a N-dimensional joint PDF representing the dynamical mass of a galaxy within N different radii, rather than single values for each radius. This approach offers advantages such as accounting for correlations between different dimensions of the output variables and enabling the estimation of uncertainties.

\subsection{Output}
\label{sec:output}

The output of our CNN is a set of N values representing the galaxy's dynamical mass enclosed within a given radius. We trained the model to reproduce the galaxy masses by minimising a mean square error (MSE) loss function:

\begin{equation}
    \label{eq:mse}
    MSE = \frac{1}{n_{data}}\sum_{i}^{n_{data}}\sum_{j}^{N} (y_{real, i,j} - y_{pred,i,j})^{2}
,\end{equation}where $y_{real, i, j}$ is the $j$-th label of the $i$-th galaxy projection in our dataset of total size $n_{data}$ and $y_{pred,i,j}$ is the neural network prediction for the same value. In our case, $y_{real, i, j}$ is the mass enclosed inside the $j$-th radius we studied for the $i$-th galaxy projection of the dataset, numerically calculated from the galaxy simulation. This process ensures the information encoded in the 32 neurons of the penultimate layer of the CNN model contains relevant information about the masses we aimed to estimate, and therefore it is suitable to use as an input for the MAF model.

The output of the MAF is a N-dimensional joint PDF, where each dimension represents the galaxy's dynamical mass enclosed within a given radius. The MAF also needs to undergo a training process so that it can learn the transformations that generate correct posterior distributions for the target data. In order to train a model that outputs a meaningful PDF ($\mathbb{P}$), we used the negative log-likelihood ($L$) of the distribution as a loss function:\\

\begin{equation}
    \label{eq:negloglik}
    L = -\ln L_{*} = \sum_{i} -\ln \left[ \mathbb{P}\left( y_{real,i} | \theta \right) \right]
,\end{equation}where $y_{real, i}$ is the set of labels of the $i$-th galaxy projection in our dataset and $\mathbb{P}$ is the PDF generated by the MAF using transformations defined by the trainable parameters $\theta$. Since the output distribution is N-dimensional, we evaluated it over a set of N labels $y_{real, i}$, which are the galaxy enclosed masses within N different radii.

\begin{figure*}[ht]
    \sidecaption
    \centering
    \includegraphics[width=12cm]{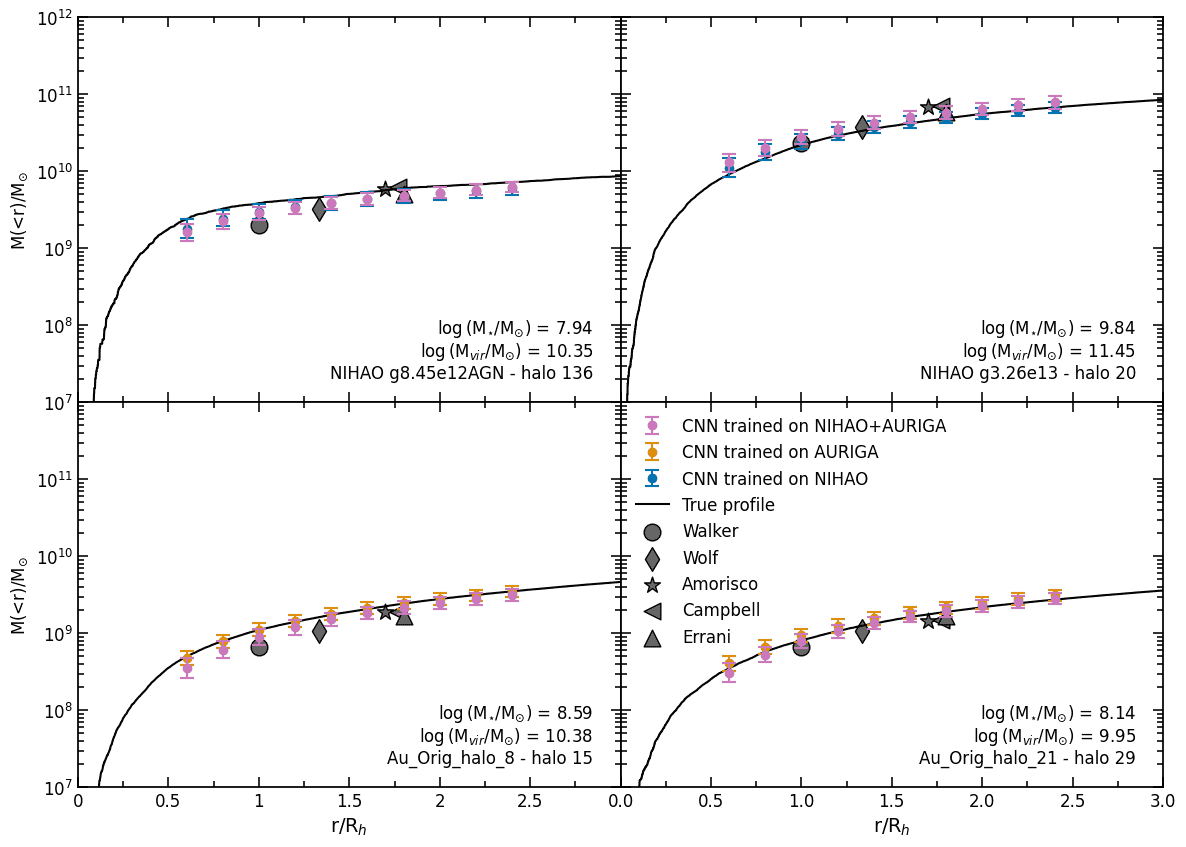}
    \caption{CNN+MAF and mass estimator predictions for four random projections of four different galaxies in the validation set. Numerical profiles of cumulative mass calculated from the simulations are plotted as solid black lines, with CNN+MAF predictions shown as circles with their corresponding 1$\sigma$ uncertainties, and mass estimator calculations displayed as grey markers with black edges. The colour of the CNN+MAF circle markers varies depending on the dataset on which the model has been trained. Each panel is labelled with the identifier of the galaxy whose results are plotted in it, specifying the simulation suite, name, and number of the halo in our halo finder. We show full posterior distributions of the CNN+MAF predictions in Fig. \protect\ref{fig:CornerPlots}.} 
\label{fig:IndividualPredictions}
\end{figure*}

\section{Results}
\label{sec:results}
\subsection{Mass estimators}
\label{sec:res-massestimators}

We present the results of applying the literature mass estimators described in Sect. \ref{sec:massestimators} to our dataset. We skipped the estimator described in \cite{Penarrubia_estim} as we find the assumption of isotropic velocities does not apply to these simulated galaxies.
    
Figure \ref{fig:full-estimators} shows the PDF of the ratio between the enclosed mass calculated by the literature mass estimators and the true enclosed mass for the galaxies in our training dataset. We also show the results of applying literature mass estimators to galaxies from each individual suite of simulations. We find all literature mass estimators to produce reasonably precise dynamical mass estimations, showing 1$\sigma$ dispersion values ranging from 0.26 for Wolf and Errani through 0.30. The accuracy of estimators also varies from one to another, and the estimators acting at the innermost radii present biases towards under-predicting the enclosed mass ($\sim$15\% for Walker and $\sim$10\% for Wolf). The trends observed are similar when analysing only AURIGA or NIHAO simulations. Nevertheless, FIRE simulations deviate slightly from the behaviour described above, exhibiting lower 1$\sigma$ dispersion and lower bias at inner radii mass estimations.

The varying performance of literature mass estimators on simulated galaxies depending on their properties or the specifics of the simulation is a topic worth exploring, but it is out of the scope of the present study. See \cite{Campbell_estim} for one such study on the Walker's and Wolf's estimators using APOSTLE simulations. Notably, the errors associated with each mass estimator in this work are generally larger than those reported in previous studies. This discrepancy arises from the application of these estimators to a diverse set of hydrodynamical simulations, as opposed to a dataset consisting solely of N-body simulations or mock data \citep[e.g.][]{Errani_estim}, or to a very uniform set of hydrodynamical simulations \citep{Campbell_estim, gonzalezsamaniego}.

\subsection{CNN}
\label{sec:res-CNN}

\begin{figure*}[ht]
    \centering
    \includegraphics[width = \textwidth]{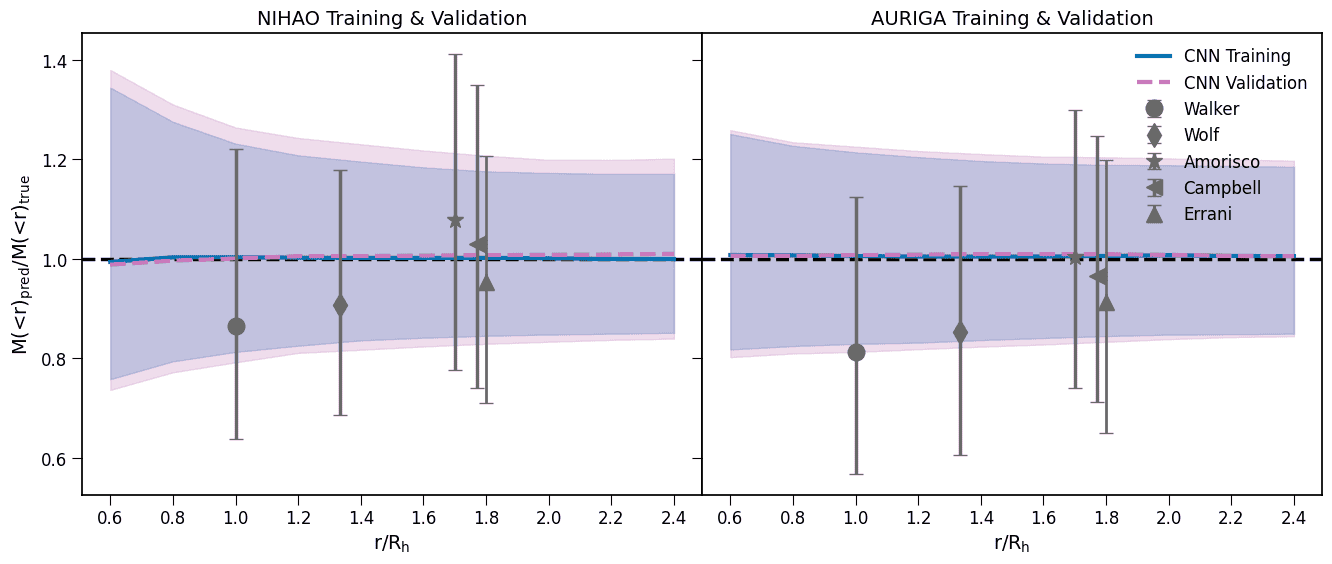}
        \caption{Ratio between the mass predicted by the CNN+MAF model and the true mass enclosed within different radii of the galaxies in the training and validation sets of NIHAO (left panel) and AURIGA (right panel) datasets. The blue line shows the median ratio determined using the mass estimated by the CNN+MAF for the training set, while the shadowed region indicates the 1$\sigma$ dispersion. The pink line and shadowed region correspond to CNN+MAF predictions for the validation set. The predicted versus true mass ratios resulting from applying literature mass estimators to either NIHAO (left panel) or AURIGA (right panel) galaxies are shown as grey symbols with 1$\sigma$ error bars.} 
        \label{fig:ratio-train-test-NH-and-AU-panels}
\end{figure*}

\begin{figure}[ht]
    \centering
    \includegraphics[width = \columnwidth]{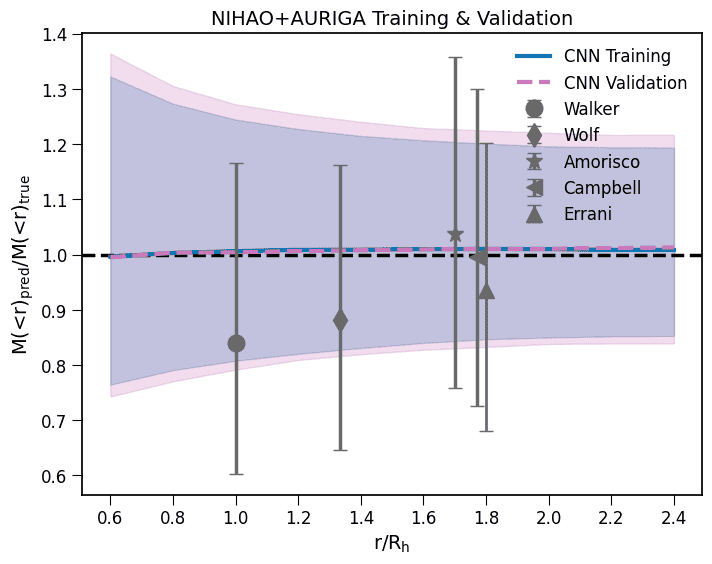}
    \caption{Ratio between the mass predicted by our CNN+MAF or the literature mass estimators and the real mass enclosed within different radii of the galaxies in the training and validation sets that combine NIHAO and AURIGA galaxies. Lines and symbols have the same meaning as in Fig. \ref{fig:ratio-train-test-NH-and-AU-panels}.}
    \label{fig:ratio-train-test}
\end{figure}

\begin{figure*}[ht]
    \centering
    \includegraphics[width = \textwidth]{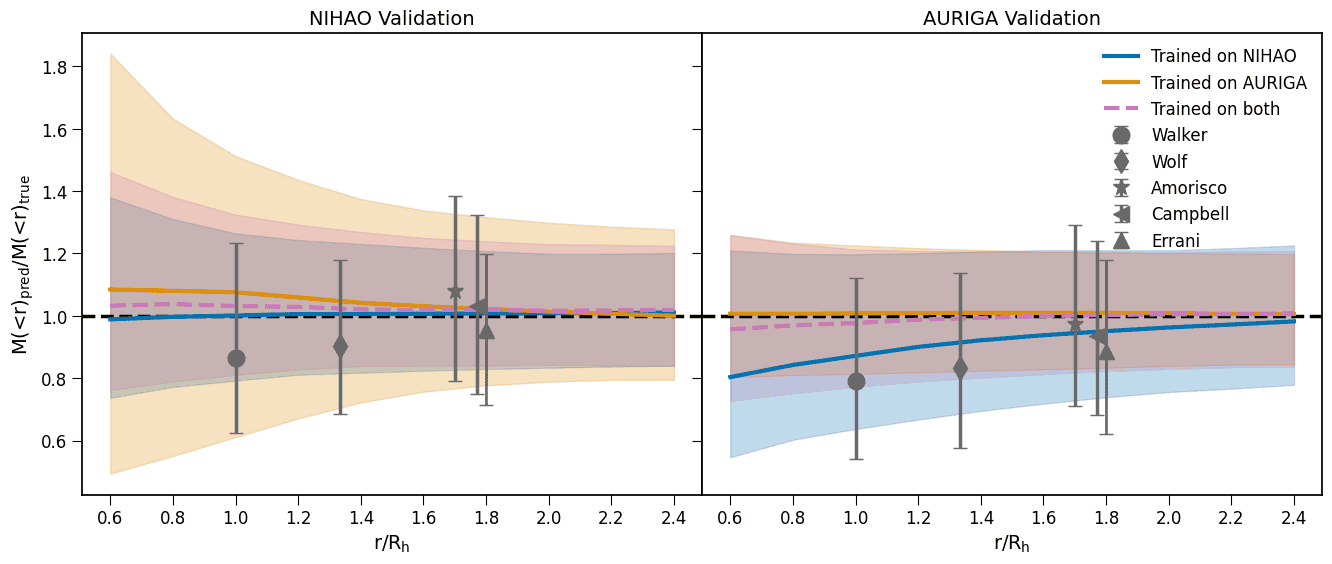}
        \caption{Ratio between the mass predicted by the CNN+MAF models, trained on different simulation suites (NIHAO-only, AURIGA-only, or a combination of the two), and the true mass enclosed within various radii of the galaxies in the NIHAO (left panel) and AURIGA (right panel) validation datasets. The blue line shows the median ratio determined using the mass estimated by the CNN+MAF trained on NIHAO galaxies, while the shadowed region indicates the 1$\sigma$. The orange line and shadowed region correspond to predictions of the CNN+MAF trained on AURIGA galaxies, and the pink line and shadowed region depict the results of the model trained on NIHAO and AURIGA galaxy projections together. The predicted versus true mass ratios resulting from applying literature mass estimators to either NIHAO (left panel) or AURIGA (right panel) galaxies are shown as grey symbols with 1$\sigma$ error bars.} 
        \label{fig:ratio-test-NH-and-AU-allmodels}
\end{figure*}

\begin{figure}[!ht]
    \centering
    \includegraphics[width = \columnwidth]{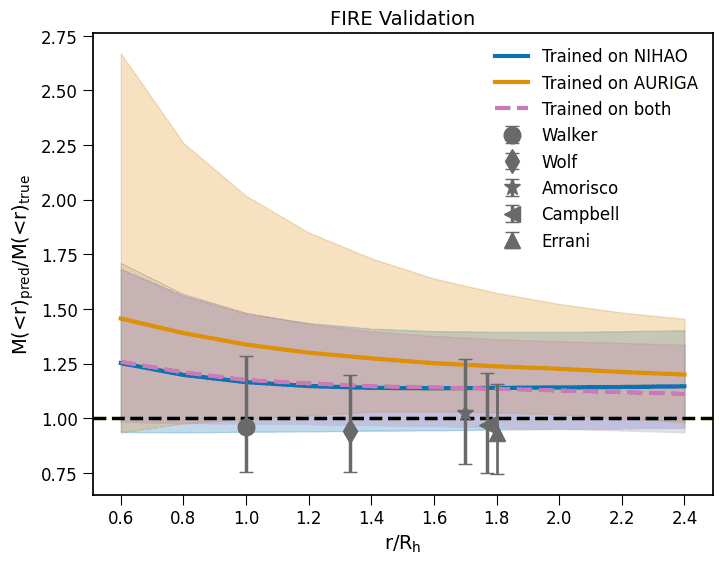}
    \caption{Ratio between the mass predicted by the CNN+MAF models, trained on different simulation suites (NIHAO-only, AURIGA-only, or a combination of the two), and the true mass enclosed within various radii of the galaxies in the FIRE dataset. Lines and symbols have the same meaning as in Fig. \ref{fig:ratio-test-NH-and-AU-allmodels}}
    \label{fig:ratio-test-FIRE-allmodels}
\end{figure}

We trained our model separately on NIHAO and AURIGA projections of galaxies, and also on a combined dataset. We carefully separated our dataset into training and validation sets, ensuring that all projections of a single galaxy are included in only one of the sets. We trained our model with a batch size of 128 using the Adam optimiser with an initial learning rate of $5\times10^{-4}$. We reduced the learning rate by a factor of 2 after 10 epochs without improvement of the validation loss, and we ended the training when the validation loss had not improved for 25 epochs straight. The training typically terminated after $\sim$100 epochs. Next we extracted the information from the penultimate layer of the CNN and used it to train the MAF model created with the python package ltu-ili \citep{Ho_2024} using 32 hidden features, 4 flow transformations, a batch size of 64, and a learning rate of $10^{-4}$.

Once the model was trained, we sampled the joint PDF obtained through our CNN+MAF model 1000 times for each galaxy projection. We obtained estimates of the dynamical enclosed mass at ten different radii by calculating the median of the resulting PDF along each dimension, and we assigned them uncertainties using the 16th and 84th percentiles. In Fig. \ref{fig:IndividualPredictions} we show example predicted mass profiles for four galaxy projections in the validation sets of NIHAO and AURIGA simulations, compared to the numerical cumulative mass profile obtained by directly analysing the simulated galaxies and to existing mass estimator results. In Fig. \ref{fig:CornerPlots} we present the full PDF obtained for each of the four galaxy projections.

Just like for the literature mass estimators, we explored the accuracy of the CNN+MAF model using the ratio between the predicted and the real mass, only we now obtained predictions at several radii on the fly using the CNN+MAF model. In the left panel of Fig. \ref{fig:ratio-train-test-NH-and-AU-panels} we show the accuracy of the model trained on NIHAO galaxies only, and we show the same for the model trained and tested on AURIGA galaxies in the right panel. We compared the model results with the accuracy of literature mass estimators in each set, and we observe that our CNN+MAF model outperforms all of them, in terms of both bias and scatter, while providing mass estimates at several radii that literature mass estimators are unable to explore. We notice that the median ratio between predicted and real enclosed mass remains constant and close to unity for both the training and validation set both for NIHAO and AURIGA galaxies. However, there are differences between the results for both simulations: our model predictions show increased scatter for NIHAO galaxies at the innermost radii, which we do not find in AURIGA. This may be a consequence of the absence of cored dark matter profiles among AURIGA galaxies, which makes dynamical mass profile shapes more uniform in comparison to NIHAO galaxies, among which we find both cusps and cores \citep{DC2014, Tollet2016, DC2017}. In both models we find more scatter on the validation set than on the training set, a hint of possible over-fitting. Constructing a mixed dataset to train on both simulations at the same time produces an intermediate result (Fig. \ref{fig:ratio-train-test}), both in terms of inner scatter and possible slight over-fitting, while the model still shows no bias and improves with respect to literature mass estimators at all radii.

Although all three models (trained on NIHAO only, on AURIGA only, and on both at the same time) outperform literature mass estimators, not all of them generalize well across all simulated galaxies. Figure \ref{fig:ratio-test-NH-and-AU-allmodels} illustrates the results of applying these models to the validation sets of NIHAO and AURIGA, respectively. We find that a model trained exclusively on one simulation set struggles to make accurate predictions for the other. In contrast, the model trained on both simulation suites together is able to predict masses accurately for each individual set.

To assess whether training on both NIHAO and AURIGA galaxies at the same time improves the model’s ability to generalise to an entirely different set of simulations, we applied the three training models to FIRE galaxies and present the results in Fig. \ref{fig:ratio-test-FIRE-allmodels}. The model trained exclusively on AURIGA galaxies exhibits systematic discrepancies when applied to the FIRE dataset, particularly at small radii. In contrast, the model trained on NIHAO galaxies demonstrates improved performance on the FIRE testing set, but still shows a tendency to overestimate masses by 20 to 25\% on average. This deviation is notably larger than that of traditional mass estimators, whose bias remains under 5\% (except 7\% for the \cite{Errani_estim} mass estimator). Despite this bias, it achieves comparable scatter relative to traditional mass estimators. We attribute the better performance of the model trained on NIHAO galaxies to the structural similarity between NIHAO and FIRE simulations, both of which produce cored dark matter density profiles, a feature not captured by AURIGA simulations. A model trained on a combined dataset of NIHAO and AURIGA galaxies further decreases the scatter compared to the NIHAO-only model, although predictions are still similarly biased.

In Table \ref{tab:MSE} we present a summary of all three models (trained exclusively on NIHAO, exclusively on AURIGA, and jointly on both simulation suites) when evaluated independently on each simulation suite. We assessed model performance using the MSE between the predicted and true values of $\log{10}$ enclosed mass at all radial bins. As a baseline for comparison, we also report the MSE obtained by assigning, to each galaxy in the validation set, the mean enclosed mass at each radius computed over all galaxies in that validation set.

As presented in Table \ref{tab:MSE}, the model trained jointly on NIHAO and AURIGA galaxies performs best on FIRE galaxies. In Fig. \ref{fig:FIRE_binned} we plot the median model prediction for 10 different bins of true values of the log$_{10}$ enclosed mass at each radius. We find the model overestimates masses most notably for the least massive galaxies, which account for a high fraction of the FIRE testing set.

\begin{table}[ht]
\caption{MSE results of different models for the log$_{10}$ enclosed mass at all radii.}
\label{tab:MSE}
\resizebox{\columnwidth}{!}{%
\begin{tabular}{c|cccc|}
\cline{2-5}
                              & \multicolumn{4}{c|}{Model}                                                                                                \\ \hline
\multicolumn{1}{|c|}{Dataset} & \multicolumn{1}{c|}{NIHAO}          & \multicolumn{1}{c|}{AURIGA}         & \multicolumn{1}{c|}{NIHAO+AURIGA}   & Baseline \\ \hline
\multicolumn{1}{|c|}{NIHAO}   & \multicolumn{1}{c|}{0.010} & \multicolumn{1}{c|}{0.032}          & \multicolumn{1}{c|}{0.011}          & 0.316   \\ \hline
\multicolumn{1}{|c|}{AURIGA}  & \multicolumn{1}{c|}{0.021}          & \multicolumn{1}{c|}{0.008} & \multicolumn{1}{c|}{0.010}          & 0.265   \\ \hline
\multicolumn{1}{|c|}{FIRE}    & \multicolumn{1}{c|}{0.016}          & \multicolumn{1}{c|}{0.034}          & \multicolumn{1}{c|}{0.015} & 0.326   \\ \hline
\end{tabular}
}
\tablefoot{
 For reference, baseline MSE values are also shown, corresponding to assigning the mean enclosed mass at each radius (computed across the validation set) to all galaxies.
}
\end{table}

\begin{figure}[ht]
        \centering
        \includegraphics[width=\columnwidth]{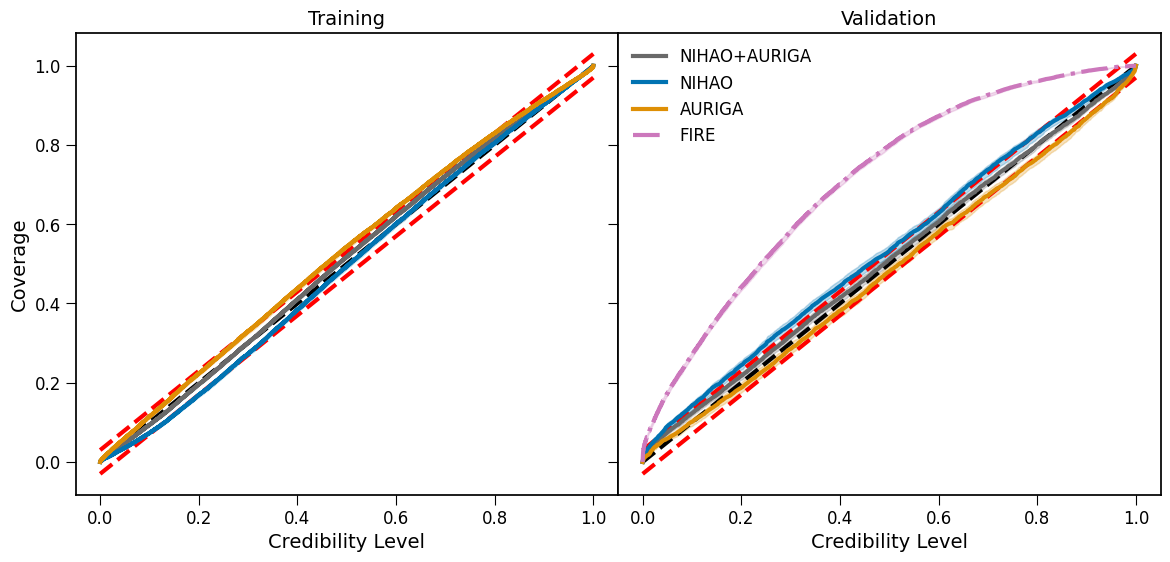}
        \caption{TARP multivariate coverage test for the training set (left panel) and validation set (right panel) of the model trained on NIHAO and AURIGA galaxies together. We show in different colours the results of the test for galaxy projections corresponding to NIHAO and AURIGA suites in each dataset. Additionally, we plot the results of the test applied to FIRE galaxies in the right panel. The dashed black lines show the expected result for a perfect calibration of the posterior, where a probability distribution contains the ground truth value within a given confidence interval with the expected frequency. Dashed red lines mark regions with a 3\% deviation from the expected statistical coverage. Solid coloured lines show the median result of the test on a subset of galaxy projections after 100 bootstrapping iterations. Shadowed regions contain the 2$\sigma$ dispersion interval.} 
        \label{fig:TARP}
    \end{figure}

As previously discussed, our model generates full posterior distributions for the enclosed masses. To evaluate the calibration of these distributions, we employed the test of accuracy with random points (TARP) methodology \citep[][]{tarp}, which assesses the statistical coverage of the posterior distributions. Statistical coverage refers to the proportion of times the true value of a parameter lies within a given credible interval of the posterior distribution. A well-calibrated posterior ensures that, for example, a 90\% credible interval contains the true value approximately 90\% of the time. Figure \ref{fig:TARP} presents the results of the TARP test for both the training and validation sets of the model trained on the combined NIHAO and AURIGA datasets. The model demonstrates well-calibrated posterior distributions for galaxies in the training set, with deviations from the expected coverage remaining below 3\% across all credibility levels. Calibration remains accurate when the model is evaluated separately on NIHAO and AURIGA galaxies within the training set, although for AURIGA galaxies, the 3\% deviation threshold is slightly exceeded at credibility levels near 50\%. The model tends to produce slightly overconfident predictions for NIHAO galaxies and more conservative estimates for AURIGA galaxies. The validation set exhibits similarly robust results, though the TARP test reveals more pronounced differences between the two simulation suites: predictions are more overconfident for AURIGA galaxies and more conservative for NIHAO galaxies.

We also applied the TARP test to the posteriors generated for FIRE galaxy projections by applying the same model. These posteriors reveal excessive overconfidence. In some cases, deviations from the expected coverage reach up to 20\%.

Further details and additional checks of the posterior calibration are provided in Appendix \ref{app:additional-checks}. In Fig. \ref{fig:perc_marginal_test} we show the results of marginal TARP tests for each dimension of the posteriors, and in Fig. \ref{fig:uncertainty-magnitude} we plot the radial profile of median posterior uncertainty.

\section{Conclusions}
\label{sec:conc}

In this work we evaluated the performance of existing literature galaxy mass estimators and constructed a novel machine-learning model able to predict the enclosed mass of dispersion-supported galaxies at any radius from line-of-sight stellar data. To do so, we used a large dataset from an extensive number of zoom-in cosmological hydrodynamical simulations from the projects NIHAO \citep{nihao_original}, AURIGA \citep{AURIGA_release}, and FIRE \citep{FIRE2_release}. 

We find that existing mass estimators work reasonably well on our dataset, producing estimates for NIHAO and AURIGA galaxies with dispersions of 0.25-0.30 on the ratio between the predicted and actual mass calculated from the simulation. The two innermost tested literature mass estimators (Walker and Wolf) systematically underestimate the enclosed mass for these galaxies by more than 10\%. We also note that the errors associated with literature mass estimators, when applied to our large set of hydrodynamical simulations, are larger than what is quoted in previous studies, likely due to such studies using a more self-similar and uniform dataset \citep{Campbell_estim, Errani_estim, gonzalezsamaniego}.

We developed a machine-learning model consisting of a CNN and a MAF that outperforms literature mass estimators when applied to simulations from the same suites it has been trained on. The model not only improves mass estimates at radii where mass estimators have been proposed in the literature (in terms of both scatter and bias), but it also provides estimates of the dynamical mass enclosed at any radii, where no mass estimators exist. Our model trained on a mixed dataset of NIHAO and AURIGA simulated galaxies has a less than 1\% bias in recovering the true enclosed galaxy  mass at any given radius, with 1$\sigma$ scatter below 21\% for radii over the projected half-light radius. A key advantage of our model is its ability to generate full posterior distributions for the enclosed masses. These posteriors are well calibrated across all credibility levels, meaning that the probability intervals (e.g. 68\% and 90\%) reliably contain the true mass values at the expected rates. Additionally, the posteriors are well calibrated for all marginals, ensuring that the 1D projections of the posterior distributions for individual parameters (e.g. mass at a specific radius) are also statistically reliable. This calibration highlights the robustness and versatility of our model for dynamical mass estimation.

We trained our model separately on galaxy projections from the NIHAO and AURIGA projects, and we find similarly good results for the two. One key difference is that the predictions of enclosed masses at inner radii show more scatter for NIHAO galaxies, which we attribute to the greater variety of inner dark matter density profile slopes in this simulation suite \citep{DC2014, Tollet2016} compared to AURIGA galaxies. We find that training exclusively on galaxies from one simulation suite and applying the model to another can lead to discrepancies in the predictions. However, when we combine NIHAO and AURIGA simulations for training and test the model on FIRE galaxies, the model captures the overall trends despite some remaining variability. Although the bias is larger compared to literature mass estimators ($\sim$20\% versus 5\%), this outcome highlights opportunities for further refining the model, for example through the application of domain adaptation techniques.

The model's lack of generalisation power might be due to different factors. Since we provide the model with the half-light radius and stellar velocity dispersion along the line of sight, and given that these properties can vary across different simulations (as shown in Fig. \ref{fig:Ms_vs_rh}), it is possible that the model learns biased relationships between these features and the enclosed mass values. Another possibility is that there is a bias in the maps used as inputs for the CNN. Biases could appear as a consequence of the varying spatial resolution of simulations, and also from a sub-optimal kernel width choice for transforming the distribution of projected stellar positions and velocities into a PDF we can sample on arrays.

At this stage, our CNN-based model can only be applied with caution to real galaxies due to the assumption that our simulation training set matches the key properties of the observed galaxies. Some discrepancies persist between hydrodynamical simulation suites and observations. For example, FIRE-2 dwarf galaxies tend to be more dispersion-dominated than their observed counterparts \citep{10.1093/mnras/sty730}, AURIGA galaxies show a systematic offset from the Faber–Jackson relation \citep{Barrientos_Acevedo_2023}, and NIHAO simulations underpredict central stellar densities \citep{Arora_2023}. Improving the applicability of our model will therefore require training and testing on a more diverse set of simulations that reproduce key observational features across a broad range of galaxy types. An additional challenge is the realistic treatment of observational effects such as uncertainties in stellar line-of-sight velocities or selection biases in stellar samples. While important, this topic is beyond the scope of our current work.

In future works, in order to improve the model's generalisability, we plan on creating a new model by following three different strategies. The first consists in changing the neural network model used from a CNN to a graph neural network \citep{GNN}, which eliminates the need to use a Gaussian kernel when creating the inputs. The next option revolves around changing the normalising flow model to one that predicts a likelihood \citep[neural likelihood estimation;][]{Alsing_2018, Alsing_2019} rather than directly outputting the posterior distribution. This would allow us to discard predictions on a new set of data never seen by the model if we find the likelihood value to be too low. Finally, another option is to apply domain adaptation techniques like domain adversarial neural networks \citep{domainadversarial}, which aim to force the model to use information that is invariant through different domains (simulation suites), improving its generalisation ability.

Our CNN+MAF method is able to estimate the mass profiles of dispersion-supported galaxies with better accuracy and with fewer assumptions than traditionally used galaxy mass estimators, at very low computational cost, on top of providing estimates of their dynamical masses across a wide range of radii.

\begin{acknowledgements}
 JSA thanks the Spanish Ministry of Economy and Competitiveness (MINECO) for support through a grant P/301404 from the Severo Ochoa project CEX2019-000920-S. CB is supported by the Spanish Ministry of Science and Innovation (MICIU/FEDER) through research grant PID2021-122603NBC22. ADC is supported by the Agencia Estatal de Investigación, under the 2023 call for Ayudas para Incentivar la Consolidación Investigadora, grant number CNS2023-144669, proyecto "TINY". The authors wish to acknowledge the contribution of the IAC High-Performance Computing support team and hardware facilities to the results of this research. The freely available software pynbody \citep{pynbody} has been used for part of this analysis. We thank the PIs of the AURIGA and FIRE-2 simulation projects for making their data publicly available at \url{https://wwwmpa.mpa-garching.mpg.de/auriga/data.html} and \url{http://flathub.flatironinstitute.org/fire}, respectively.

\end{acknowledgements}

\bibliographystyle{aa}
\bibliography{bib.bib} 

\begin{thebibliography}{48}
\expandafter\ifx\csname natexlab\endcsname\relax\def\natexlab#1{#1}\fi

\bibitem[{Alsing {et~al.}(2019)Alsing, Charnock, Feeney, \& Wandelt}]{Alsing_2019}
Alsing, J., Charnock, T., Feeney, S., \& Wandelt, B. 2019, \mnras

\bibitem[{Alsing {et~al.}(2018)Alsing, Wandelt, \& Feeney}]{Alsing_2018}
Alsing, J., Wandelt, B., \& Feeney, S. 2018, \mnras, 477, 2874–2885

\bibitem[{{Amorisco} \& {Evans}(2012)}]{Amorisco_estim}
{Amorisco}, N.~C. \& {Evans}, N.~W. 2012, \mnras, 419, 184

\bibitem[{Arora {et~al.}(2023)Arora, Courteau, Stone, \& Macciò}]{Arora_2023}
Arora, N., Courteau, S., Stone, C., \& Macciò, A.~V. 2023, \mnras, 522, 1208–1227

\bibitem[{Barrientos Acevedo {et~al.}(2023)Barrientos Acevedo, van der Wel, Baes, Grand, Kapoor, Camps, de Graaff, Straatman, \& Bezanson}]{Barrientos_Acevedo_2023}
Barrientos Acevedo, D., van der Wel, A., Baes, M., {et~al.} 2023, \mnras, 524, 907–922

\bibitem[{Blank {et~al.}(2019)Blank, Macciò, Dutton, \& Obreja}]{nihao_bh}
Blank, M., Macciò, A.~V., Dutton, A.~A., \& Obreja, A. 2019, \mnras, 487, 5476–5489

\bibitem[{{Brook} {et~al.}(2012){Brook}, {Stinson}, {Gibson}, {Wadsley}, \& {Quinn}}]{brook12b}
{Brook}, C.~B., {Stinson}, G., {Gibson}, B.~K., {Wadsley}, J., \& {Quinn}, T. 2012, \mnras, 424, 1275

\bibitem[{Buck {et~al.}(2019)Buck, Obreja, Macciò, Minchev, Dutton, \& Ostriker}]{NIHAO_UHD}
Buck, T., Obreja, A., Macciò, A.~V., {et~al.} 2019, \mnras, 491, 3461

\bibitem[{{Campbell} {et~al.}(2017){Campbell}, {Frenk}, {Jenkins}, {Eke}, {Navarro}, {Sawala}, {Schaller}, {Fattahi}, {Oman}, \& {Theuns}}]{Campbell_estim}
{Campbell}, D. J.~R., {Frenk}, C.~S., {Jenkins}, A., {et~al.} 2017, \mnras, 469, 2335

\bibitem[{{Chabrier}(2003)}]{Chabrier03}
{Chabrier}, G. 2003, \pasp, 115, 763

\bibitem[{{Correa} {et~al.}(2017){Correa}, {Schaye}, {Clauwens}, {Bower}, {Crain}, {Schaller}, {Theuns}, \& {Thob}}]{kappa_salva}
{Correa}, C.~A., {Schaye}, J., {Clauwens}, B., {et~al.} 2017, \mnras, 472, L45

\bibitem[{de los Rios {et~al.}(2023)de los Rios, Petač, Zaldivar, Bonaventura, Calore, \& Iocco}]{fabio_iocco}
de los Rios, M., Petač, M., Zaldivar, B., {et~al.} 2023, \mnras, 525, 6015

\bibitem[{{Di Cintio} {et~al.}(2017){Di Cintio}, {Brook}, {Dutton}, {Macci{\`o}}, {Obreja}, \& {Dekel}}]{DC2017}
{Di Cintio}, A., {Brook}, C.~B., {Dutton}, A.~A., {et~al.} 2017, \mnras, 466, L1

\bibitem[{{Di Cintio} {et~al.}(2014){Di Cintio}, {Brook}, {Macci{\`o}}, {Stinson}, {Knebe}, {Dutton}, \& {Wadsley}}]{DC2014}
{Di Cintio}, A., {Brook}, C.~B., {Macci{\`o}}, A.~V., {et~al.} 2014, \mnras, 437, 415

\bibitem[{Dutton {et~al.}(2020)Dutton, Buck, Macciò, Dixon, Blank, \& Obreja}]{nihao_nth}
Dutton, A.~A., Buck, T., Macciò, A.~V., {et~al.} 2020, \mnras, 499, 2648

\bibitem[{El-Badry {et~al.}(2018)El-Badry, Bradford, Quataert, Geha, Boylan-Kolchin, Weisz, Wetzel, Hopkins, Chan, Fitts, Kereš, \& Faucher-Giguère}]{10.1093/mnras/sty730}
El-Badry, K., Bradford, J., Quataert, E., {et~al.} 2018, \mnras, 477, 1536

\bibitem[{{Errani} {et~al.}(2018){Errani}, {Pe{\~n}arrubia}, \& {Walker}}]{Errani_estim}
{Errani}, R., {Pe{\~n}arrubia}, J., \& {Walker}, M.~G. 2018, \mnras, 481, 5073

\bibitem[{{Exp{\'o}sito-M{\'a}rquez} {et~al.}(2023){Exp{\'o}sito-M{\'a}rquez}, {Brook}, {Huertas-Company}, {Di Cintio}, {Macci{\`o}}, {Grand}, {Battaglia}, \& {Arjona-G{\'a}lvez}}]{julen}
{Exp{\'o}sito-M{\'a}rquez}, J., {Brook}, C.~B., {Huertas-Company}, M., {et~al.} 2023, \mnras, 519, 4384

\bibitem[{Ganin {et~al.}(2016)Ganin, Ustinova, Ajakan, Germain, Larochelle, Laviolette, Marchand, \& Lempitsky}]{domainadversarial}
Ganin, Y., Ustinova, E., Ajakan, H., {et~al.} 2016, Domain-Adversarial Training of Neural Networks

\bibitem[{{Gonz{\'a}lez-Samaniego} {et~al.}(2017){Gonz{\'a}lez-Samaniego}, {Bullock}, {Boylan-Kolchin}, {Fitts}, {Elbert}, {Hopkins}, {Kere{\v{s}}}, \& {Faucher-Gigu{\`e}re}}]{gonzalezsamaniego}
{Gonz{\'a}lez-Samaniego}, A., {Bullock}, J.~S., {Boylan-Kolchin}, M., {et~al.} 2017, \mnras, 472, 4786

\bibitem[{Grand {et~al.}(2024)Grand, Fragkoudi, Gómez, Jenkins, Marinacci, Pakmor, \& Springel}]{AURIGA_release}
Grand, R. J.~J., Fragkoudi, F., Gómez, F.~A., {et~al.} 2024, \mnras, 532, 1814

\bibitem[{{Grand} {et~al.}(2017){Grand}, {G{\'o}mez}, {Marinacci}, {Pakmor}, {Springel}, {Campbell}, {Frenk}, {Jenkins}, \& {White}}]{2017auriga}
{Grand}, R. J.~J., {G{\'o}mez}, F.~A., {Marinacci}, F., {et~al.} 2017, \mnras, 467, 179

\bibitem[{Ho {et~al.}(2024{\natexlab{a}})Ho, Bartlett, Chartier, Cuesta-Lazaro, Ding, Lapel, Lemos, Lovell, Makinen, Modi, Pandya, Pandey, Perez, Wandelt, \& Bryan}]{ltuili}
Ho, M., Bartlett, D.~J., Chartier, N., {et~al.} 2024{\natexlab{a}}, LtU-ILI: An All-in-One Framework for Implicit Inference in Astrophysics and Cosmology

\bibitem[{Ho {et~al.}(2024{\natexlab{b}})Ho, Bartlett, Chartier, Cuesta-Lazaro, Ding, Lapel, Lemos, Lovell, Makinen, Modi, Pandya, Pandey, Perez, Wandelt, \& Bryan}]{Ho_2024}
Ho, M., Bartlett, D.~J., Chartier, N., {et~al.} 2024{\natexlab{b}}, OJAp, 7

\bibitem[{Ho {et~al.}(2019)Ho, Rau, Ntampaka, Farahi, Trac, \& Póczos}]{Ho_2019_clusters}
Ho, M., Rau, M.~M., Ntampaka, M., {et~al.} 2019, \apj, 887, 25

\bibitem[{Hopkins {et~al.}(2018{\natexlab{a}})Hopkins, Wetzel, Kereš, Faucher-Giguère, Quataert, Boylan-Kolchin, Murray, Hayward, Garrison-Kimmel, Hummels, Feldmann, Torrey, Ma, Anglés-Alcázar, Su, Orr, Schmitz, Escala, Sanderson, Grudić, Hafen, Kim, Fitts, Bullock, Wheeler, Chan, Elbert, \& Narayanan}]{Hopkins_2018}
Hopkins, P.~F., Wetzel, A., Kereš, D., {et~al.} 2018{\natexlab{a}}, \mnras, 480, 800–863

\bibitem[{Hopkins {et~al.}(2018{\natexlab{b}})Hopkins, Wetzel, Kereš, Faucher-Giguère, Quataert, Boylan-Kolchin, Murray, Hayward, Garrison-Kimmel, Hummels, Feldmann, Torrey, Ma, Anglés-Alcázar, Su, Orr, Schmitz, Escala, Sanderson, Grudić, Hafen, Kim, Fitts, Bullock, Wheeler, Chan, Elbert, \& Narayanan}]{fire}
Hopkins, P.~F., Wetzel, A., Kereš, D., {et~al.} 2018{\natexlab{b}}, \mnras, 480, 800–863

\bibitem[{{Kim} {et~al.}(2014){Kim}, {Abel}, {Agertz}, {Bryan}, {Ceverino}, {Christensen}, {Conroy}, {Dekel}, {Gnedin}, {Goldbaum}, {Guedes}, {Hahn}, {Hobbs}, {Hopkins}, {Hummels}, {Iannuzzi}, {Keres}, {Klypin}, {Kravtsov}, {Krumholz}, {Kuhlen}, {Leitner}, {Madau}, {Mayer}, {Moody}, {Nagamine}, {Norman}, {Onorbe}, {O'Shea}, {Pillepich}, {Primack}, {Quinn}, {Read}, {Robertson}, {Rocha}, {Rudd}, {Shen}, {Smith}, {Szalay}, {Teyssier}, {Thompson}, {Todoroki}, {Turk}, {Wadsley}, {Wise}, {Zolotov}, \& {AGORA Collaboration29}}]{AGORA_cosmo}
{Kim}, J.-h., {Abel}, T., {Agertz}, O., {et~al.} 2014, \apjs, 210, 14

\bibitem[{Lemos {et~al.}(2023)Lemos, Coogan, Hezaveh, \& Perreault-Levasseur}]{tarp}
Lemos, P., Coogan, A., Hezaveh, Y., \& Perreault-Levasseur, L. 2023, Sampling-Based Accuracy Testing of Posterior Estimators for General Inference

\bibitem[{Navarro {et~al.}(1997)Navarro, Frenk, \& White}]{NFW}
Navarro, J.~F., Frenk, C.~S., \& White, S. D.~M. 1997, \apj, 490, 493–508

\bibitem[{Nguyen {et~al.}(2023)Nguyen, Mishra-Sharma, Williams, \& Necib}]{Nguyen_2023}
Nguyen, T., Mishra-Sharma, S., Williams, R., \& Necib, L. 2023, Phys. Rev. D, 107

\bibitem[{Papamakarios {et~al.}(2021)Papamakarios, Nalisnick, Rezende, Mohamed, \& Lakshminarayanan}]{normflows}
Papamakarios, G., Nalisnick, E., Rezende, D.~J., Mohamed, S., \& Lakshminarayanan, B. 2021, JMLR, 22, 1

\bibitem[{Papamakarios {et~al.}(2018)Papamakarios, Pavlakou, \& Murray}]{maf}
Papamakarios, G., Pavlakou, T., \& Murray, I. 2018, Masked Autoregressive Flow for Density Estimation

\bibitem[{Paszke {et~al.}(2019)Paszke, Gross, Massa, Lerer, Bradbury, Chanan, Killeen, Lin, Gimelshein, Antiga, Desmaison, Köpf, Yang, DeVito, Raison, Tejani, Chilamkurthy, Steiner, Fang, Bai, \& Chintala}]{paszke2019pytorch}
Paszke, A., Gross, S., Massa, F., {et~al.} 2019, PyTorch: An Imperative Style, High-Performance Deep Learning Library

\bibitem[{{Pe{\~n}arrubia} {et~al.}(2008){Pe{\~n}arrubia}, {McConnachie}, \& {Navarro}}]{Penarrubia_estim}
{Pe{\~n}arrubia}, J., {McConnachie}, A.~W., \& {Navarro}, J.~F. 2008, \apj, 672, 904

\bibitem[{{Planck Collaboration} {et~al.}(2014){Planck Collaboration}, {Ade}, {Aghanim}, {Armitage-Caplan}, {Arnaud}, {Ashdown}, {Atrio-Barandela}, {Aumont}, {Baccigalupi}, {Banday}, {Barreiro}, {Bartlett}, {Battaner}, {Benabed}, {Beno{\^\i}t}, {Benoit-L{\'e}vy}, {Bernard}, {Bersanelli}, {Bielewicz}, {Bobin}, {Bock}, {Bonaldi}, {Bond}, {Borrill}, {Bouchet}, {Bridges}, {Bucher}, {Burigana}, {Butler}, {Calabrese}, {Cappellini}, {Cardoso}, {Catalano}, {Challinor}, {Chamballu}, {Chary}, {Chen}, {Chiang}, {Chiang}, {Christensen}, {Church}, {Clements}, {Colombi}, {Colombo}, {Couchot}, {Coulais}, {Crill}, {Curto}, {Cuttaia}, {Danese}, {Davies}, {Davis}, {de Bernardis}, {de Rosa}, {de Zotti}, {Delabrouille}, {Delouis}, {D{\'e}sert}, {Dickinson}, {Diego}, {Dolag}, {Dole}, {Donzelli}, {Dor{\'e}}, {Douspis}, {Dunkley}, {Dupac}, {Efstathiou}, {Elsner}, {En{\ss}lin}, {Eriksen}, {Finelli}, {Forni}, {Frailis}, {Fraisse}, {Franceschi}, {Gaier}, {Galeotta}, {Galli}, {Ganga}, {Giard}, {Giardino}, {Giraud-H{\'e}raud},
  {Gjerl{\o}w}, {Gonz{\'a}lez-Nuevo}, {G{\'o}rski}, {Gratton}, {Gregorio}, {Gruppuso}, {Gudmundsson}, {Haissinski}, {Hamann}, {Hansen}, {Hanson}, {Harrison}, {Henrot-Versill{\'e}}, {Hern{\'a}ndez-Monteagudo}, {Herranz}, {Hildebrandt}, {Hivon}, {Hobson}, {Holmes}, {Hornstrup}, {Hou}, {Hovest}, {Huffenberger}, {Jaffe}, {Jaffe}, {Jewell}, {Jones}, {Juvela}, {Keih{\"a}nen}, {Keskitalo}, {Kisner}, {Kneissl}, {Knoche}, {Knox}, {Kunz}, {Kurki-Suonio}, {Lagache}, {L{\"a}hteenm{\"a}ki}, {Lamarre}, {Lasenby}, {Lattanzi}, {Laureijs}, {Lawrence}, {Leach}, {Leahy}, {Leonardi}, {Le{\'o}n-Tavares}, {Lesgourgues}, {Lewis}, {Liguori}, {Lilje}, {Linden-V{\o}rnle}, {L{\'o}pez-Caniego}, {Lubin}, {Mac{\'\i}as-P{\'e}rez}, {Maffei}, {Maino}, {Mandolesi}, {Maris}, {Marshall}, {Martin}, {Mart{\'\i}nez-Gonz{\'a}lez}, {Masi}, {Massardi}, {Matarrese}, {Matthai}, {Mazzotta}, {Meinhold}, {Melchiorri}, {Melin}, {Mendes}, {Menegoni}, {Mennella}, {Migliaccio}, {Millea}, {Mitra}, {Miville-Desch{\^e}nes}, {Moneti}, {Montier}, {Morgante},
  {Mortlock}, {Moss}, {Munshi}, {Murphy}, {Naselsky}, {Nati}, {Natoli}, {Netterfield}, {N{\o}rgaard-Nielsen}, {Noviello}, {Novikov}, {Novikov}, {O'Dwyer}, {Osborne}, {Oxborrow}, {Paci}, {Pagano}, {Pajot}, {Paladini}, {Paoletti}, {Partridge}, {Pasian}, {Patanchon}, {Pearson}, {Pearson}, {Peiris}, {Perdereau}, {Perotto}, {Perrotta}, {Pettorino}, {Piacentini}, {Piat}, {Pierpaoli}, {Pietrobon}, {Plaszczynski}, {Platania}, {Pointecouteau}, {Polenta}, {Ponthieu}, {Popa}, {Poutanen}, {Pratt}, {Pr{\'e}zeau}, {Prunet}, {Puget}, {Rachen}, {Reach}, {Rebolo}, {Reinecke}, {Remazeilles}, {Renault}, {Ricciardi}, {Riller}, {Ristorcelli}, {Rocha}, {Rosset}, {Roudier}, {Rowan-Robinson}, {Rubi{\~n}o-Mart{\'\i}n}, {Rusholme}, {Sandri}, {Santos}, {Savelainen}, {Savini}, {Scott}, {Seiffert}, {Shellard}, {Spencer}, {Starck}, {Stolyarov}, {Stompor}, {Sudiwala}, {Sunyaev}, {Sureau}, {Sutton}, {Suur-Uski}, {Sygnet}, {Tauber}, {Tavagnacco}, {Terenzi}, {Toffolatti}, {Tomasi}, {Tristram}, {Tucci}, {Tuovinen}, {T{\"u}rler}, {Umana},
  {Valenziano}, {Valiviita}, {Van Tent}, {Vielva}, {Villa}, {Vittorio}, {Wade}, {Wandelt}, {Wehus}, {White}, {White}, {Wilkinson}, {Yvon}, {Zacchei}, \& {Zonca}}]{planckcosmo}
{Planck Collaboration}, {Ade}, P.~A.~R., {Aghanim}, N., {et~al.} 2014, \aap, 571, A16

\bibitem[{{Pontzen} {et~al.}(2013){Pontzen}, {Ro{\v s}kar}, {Stinson}, {Woods}, {Reed}, {Coles}, \& {Quinn}}]{pynbody}
{Pontzen}, A., {Ro{\v s}kar}, R., {Stinson}, G.~S., {et~al.} 2013, {pynbody: Astrophysics Simulation Analysis for Python}, astrophysics Source Code Library, ascl:1305.002

\bibitem[{Sawala {et~al.}(2016)Sawala, Frenk, Fattahi, Navarro, Bower, Crain, Vecchia, Furlong, Helly, Jenkins, Oman, Schaller, Schaye, Theuns, Trayford, \& White}]{apostle}
Sawala, T., Frenk, C.~S., Fattahi, A., {et~al.} 2016, \mnras, 457, 1931

\bibitem[{Scarselli {et~al.}(2009)Scarselli, Gori, Tsoi, Hagenbuchner, \& Monfardini}]{GNN}
Scarselli, F., Gori, M., Tsoi, A.~C., Hagenbuchner, M., \& Monfardini, G. 2009, IEEE Trans. Neural Netw., 20, 61

\bibitem[{{Schmidt}(1959)}]{KS}
{Schmidt}, M. 1959, \apj, 129, 243

\bibitem[{Shen {et~al.}(2010)Shen, Wadsley, \& Stinson}]{shen}
Shen, S., Wadsley, J., \& Stinson, G. 2010, \mnras, 407, 1581–1596

\bibitem[{{Stinson} {et~al.}(2006){Stinson}, {Seth}, {Katz}, {Wadsley}, {Governato}, \& {Quinn}}]{stinson06}
{Stinson}, G., {Seth}, A., {Katz}, N., {et~al.} 2006, \mnras, 373, 1074

\bibitem[{{Stinson} {et~al.}(2013){Stinson}, {Brook}, {Macci{\`o}}, {Wadsley}, {Quinn}, \& {Couchman}}]{stinson13}
{Stinson}, G.~S., {Brook}, C., {Macci{\`o}}, A.~V., {et~al.} 2013, \mnras, 428, 129

\bibitem[{{Tollet} {et~al.}(2016){Tollet}, {Macci{\`o}}, {Dutton}, {Stinson}, {Wang}, {Penzo}, {Gutcke}, {Buck}, {Kang}, {Brook}, {Di Cintio}, {Keller}, \& {Wadsley}}]{Tollet2016}
{Tollet}, E., {Macci{\`o}}, A.~V., {Dutton}, A.~A., {et~al.} 2016, \mnras, 456, 3542

\bibitem[{{Walker} {et~al.}(2009){Walker}, {Mateo}, {Olszewski}, {Pe{\~n}arrubia}, {Evans}, \& {Gilmore}}]{Walker_estim}
{Walker}, M.~G., {Mateo}, M., {Olszewski}, E.~W., {et~al.} 2009, \apj, 704, 1274

\bibitem[{Wang {et~al.}(2015)Wang, Dutton, Stinson, Macciò, Penzo, Kang, Keller, \& Wadsley}]{nihao_original}
Wang, L., Dutton, A.~A., Stinson, G.~S., {et~al.} 2015, \mnras, 454, 83–94

\bibitem[{Wetzel {et~al.}(2023)Wetzel, Hayward, Sanderson, Ma, Anglés-Alcázar, Feldmann, Chan, El-Badry, Wheeler, Garrison-Kimmel, Nikakhtar, Panithanpaisal, Arora, Gurvich, Samuel, Sameie, Pandya, Hafen, Hummels, Loebman, Boylan-Kolchin, Bullock, Faucher-Giguère, Kereš, Quataert, \& Hopkins}]{FIRE2_release}
Wetzel, A., Hayward, C.~C., Sanderson, R.~E., {et~al.} 2023, \apjs, 265, 44

\bibitem[{{Wolf} {et~al.}(2010){Wolf}, {Martinez}, {Bullock}, {Kaplinghat}, {Geha}, {Mu{\~n}oz}, {Simon}, \& {Avedo}}]{Wolf_estim}
{Wolf}, J., {Martinez}, G.~D., {Bullock}, J.~S., {et~al.} 2010, \mnras, 406, 1220

\end{thebibliography}

\begin{appendix}
    \section{Bivariate kernel density estimation}
    \label{app:kde}

    Kernel density estimation is a non-parametric technique used to estimate the PDF of a random variable. Bivariate KDE extends this concept to two dimensions, estimating the joint PDF of two random variables.

KDE estimates the value of the underlying PDF at the point x according to

    \begin{equation}
        f_{h}(x) = \frac{1}{n\left|\mathbb{H} \right|^{1/2}}\sum_{i=1}^{n}K\left[ \mathbb{H}^{-1/2}\left( x - \mathbf{X_{i}}\right)\right]
    ,\end{equation}where $\mathbf{X_{1}},$ $\mathbf{X_{2}},$ ..., $\mathbf{X_{n}}$ constitute a sample of size n of the random variable(s) whose PDF we aim to estimate. In the case of bivariate KDE, $\mathbf{X_{i}}$ represents two values $\mathbf{X_{i}} = \left( \mathbf{X_{i,1}}, \mathbf{X_{i,2}}\right)$, which are joint samples of the two random variables for which we wanted to estimate the PDF. $K$ is a kernel function and $\mathbb{H}$ is a bandwidth matrix whose dimensions match those of $\mathbf{X_{i}}$ (in this case, 2x2).

    Effectively, the KDE evaluates the PDF at $x$ by adding the contributions of each point $\mathbf{X_{i}}$, weighted by the kernel function $K$, based on the distance between $x$ and $\mathbf{X_{i}}$, scaled by the bandwidth matrix $\mathbb{H}$. We used a 2D Gaussian kernel that follows

    \begin{equation}
        K(\mathbf{u}) = (2\pi)^{-3/2} \left| \mathbb{H} \right|^{1/2} \exp\left(-\frac{1}{2} \mathbf{u}^{T} \mathbb{H}\mathbf{u}\right)
    ,\end{equation}where $\mathbf{u}=x-\mathbf{X_{i}}$. The bandwidth matrix $\mathbb{H}$ is calculated by multiplying the covariance matrix of the data by a factor of $n^{-1/6}$. Typically, n is the full size of the sample. However, we redefined it as the number of stars inside half the half-light radius ($0.5 \times R_{\rm h}$).

    \section{MAF posterior corner plots}

\begin{figure*}[ht]
        \centering
        \begin{subfigure}{0.49\textwidth}
            \centering
            \includegraphics[width=\textwidth]{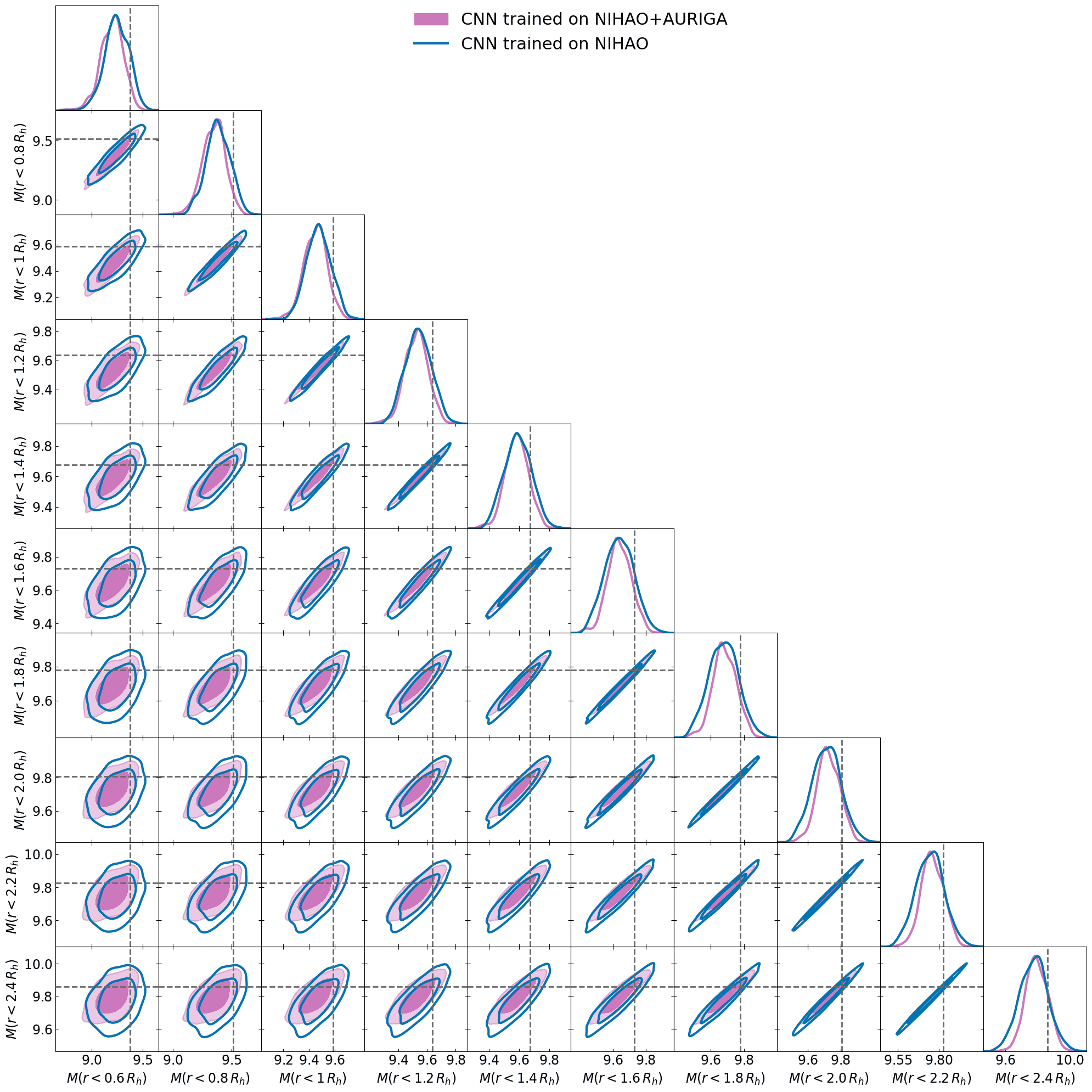}
            \caption{\small }    
            \label{fig:profile1contours}
        \end{subfigure}
        \hfill
        \begin{subfigure}{0.49\textwidth}  
            \centering 
            \includegraphics[width=\textwidth]{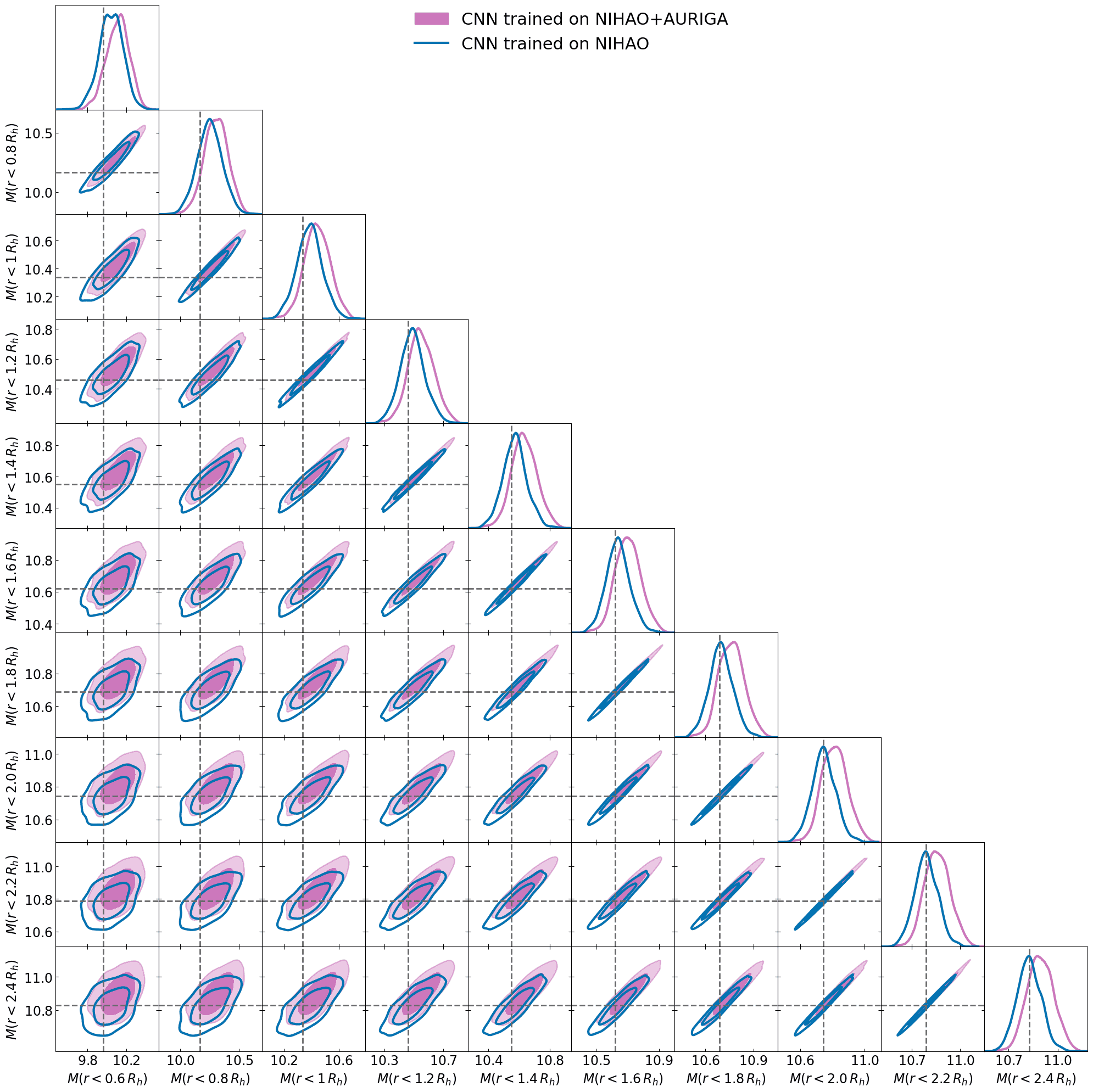}
            \caption{\small }    
            \label{fig:profile2contours}
        \end{subfigure}
        \vskip\baselineskip
        \begin{subfigure}{0.49\textwidth}   
            \centering 
            \includegraphics[width=\textwidth]{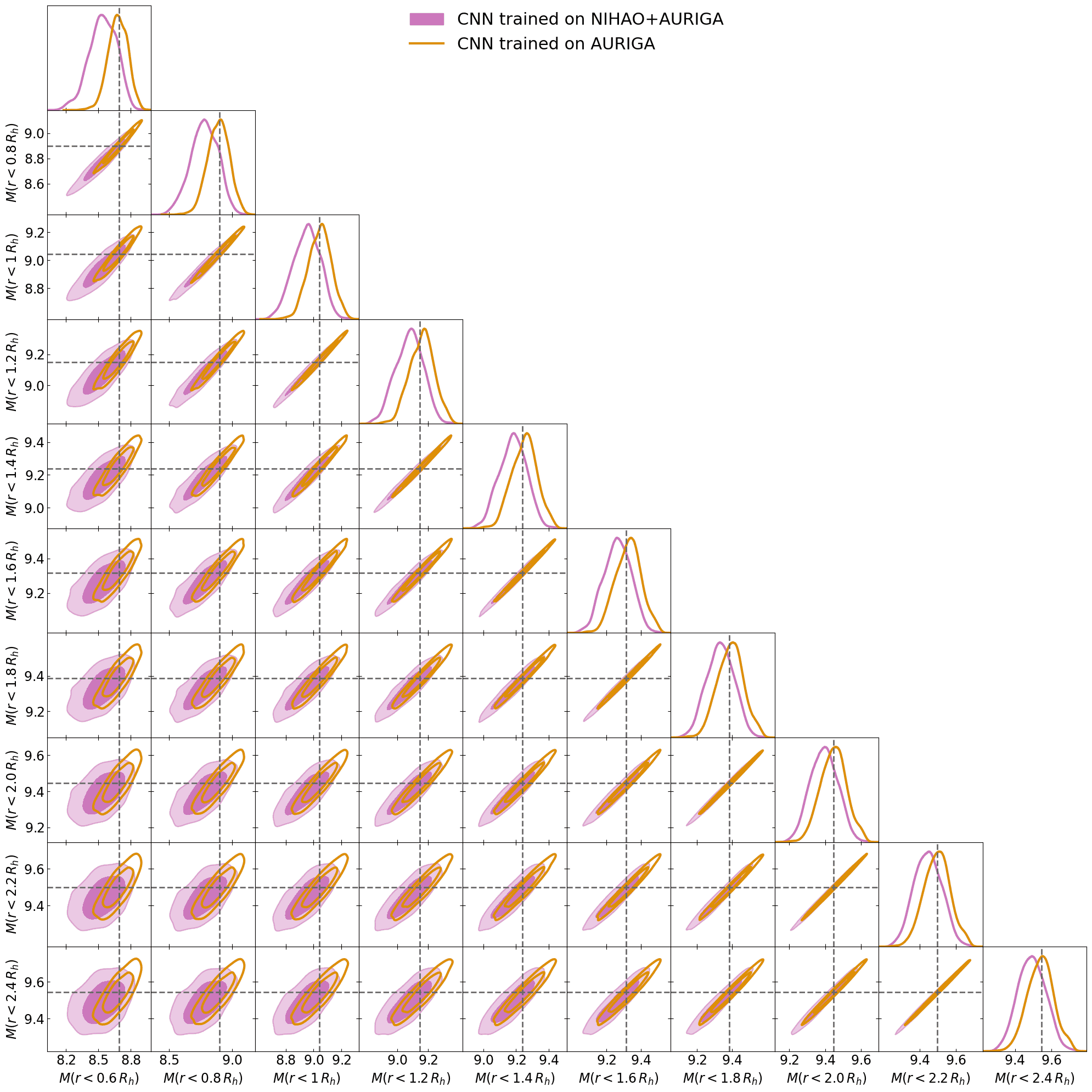}
            \caption{\small }    
            \label{fig:mprofile3contours}
        \end{subfigure}
        \hfill
        \begin{subfigure}{0.49\textwidth}   
            \centering 
            \includegraphics[width=\textwidth]{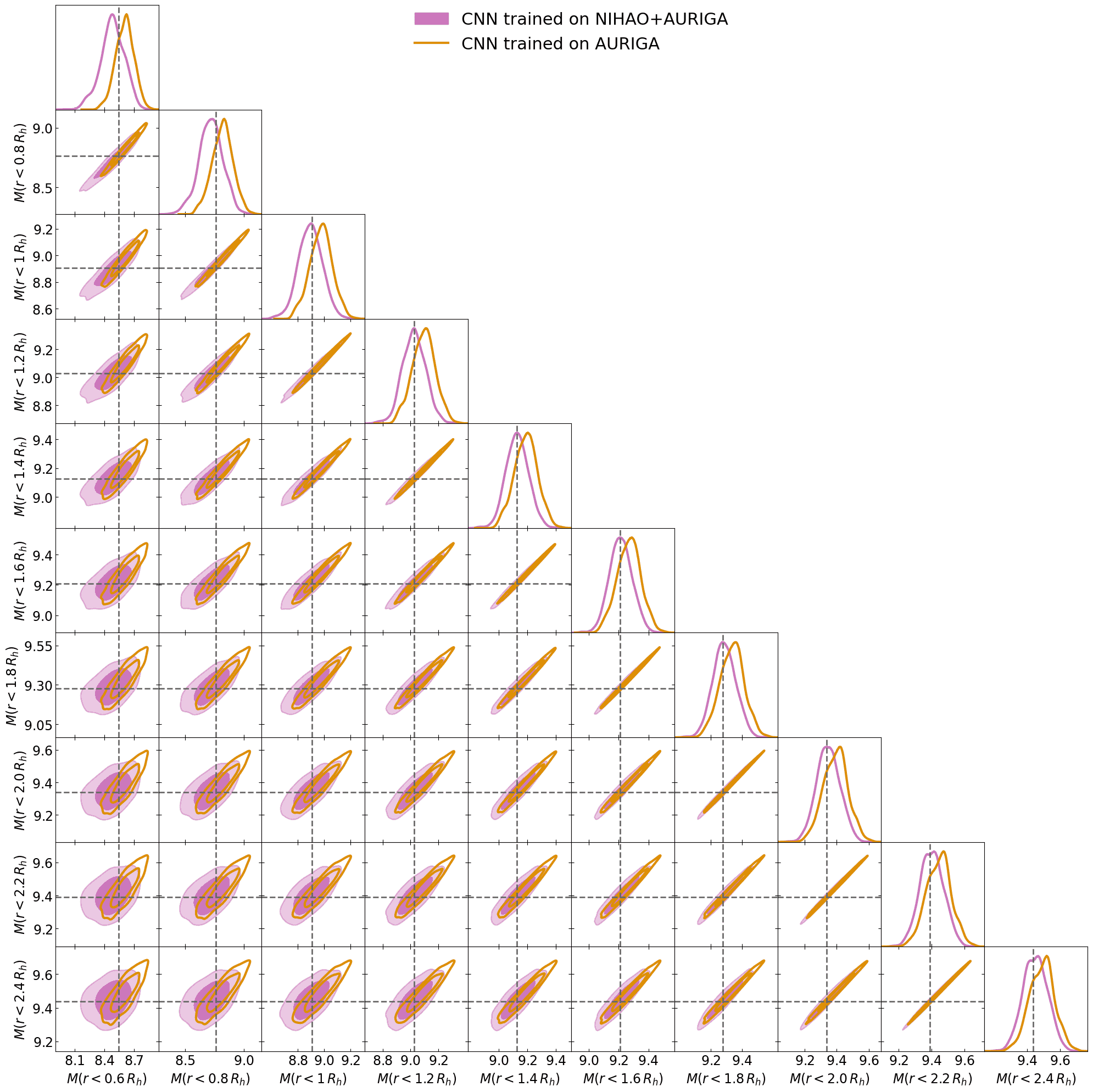}
            \caption{\small }    
            \label{fig:profile4contours}
        \end{subfigure}
        \caption{PDFs generated by sampling 1000 times the MAF model for the four galaxy projections in the validation set shown in Fig. \protect\ref{fig:IndividualPredictions}: NIHAO g8.45e12AGN - halo 136 (a), NIHAO g3.26e13 - halo 20 (b), AURIGA Original\_halo\_8 - halo 15 (c), and AURIGA Original\_halo\_21 - halo 29 (d). Contours enclose 68\% and 95\% of the samples. The colour of the contours varies according to the dataset on which the model has been trained. Dashed grey lines mark the true values of the enclosed mass calculated directly from the simulation.} 
        \label{fig:CornerPlots}
    \end{figure*}

    In Fig. \ref{fig:CornerPlots} we present pairwise contour plots of the posterior samples generated by the CNN+MAF models for the four validation galaxies shown in Fig. \ref{fig:IndividualPredictions}. These plots illustrate the joint distributions of the inferred values for the enclosed mass at all radii of interest.

    \clearpage
    \section{Predictions for individual radial bins}
\begin{figure*}
    \centering
    \includegraphics[width=\textwidth]{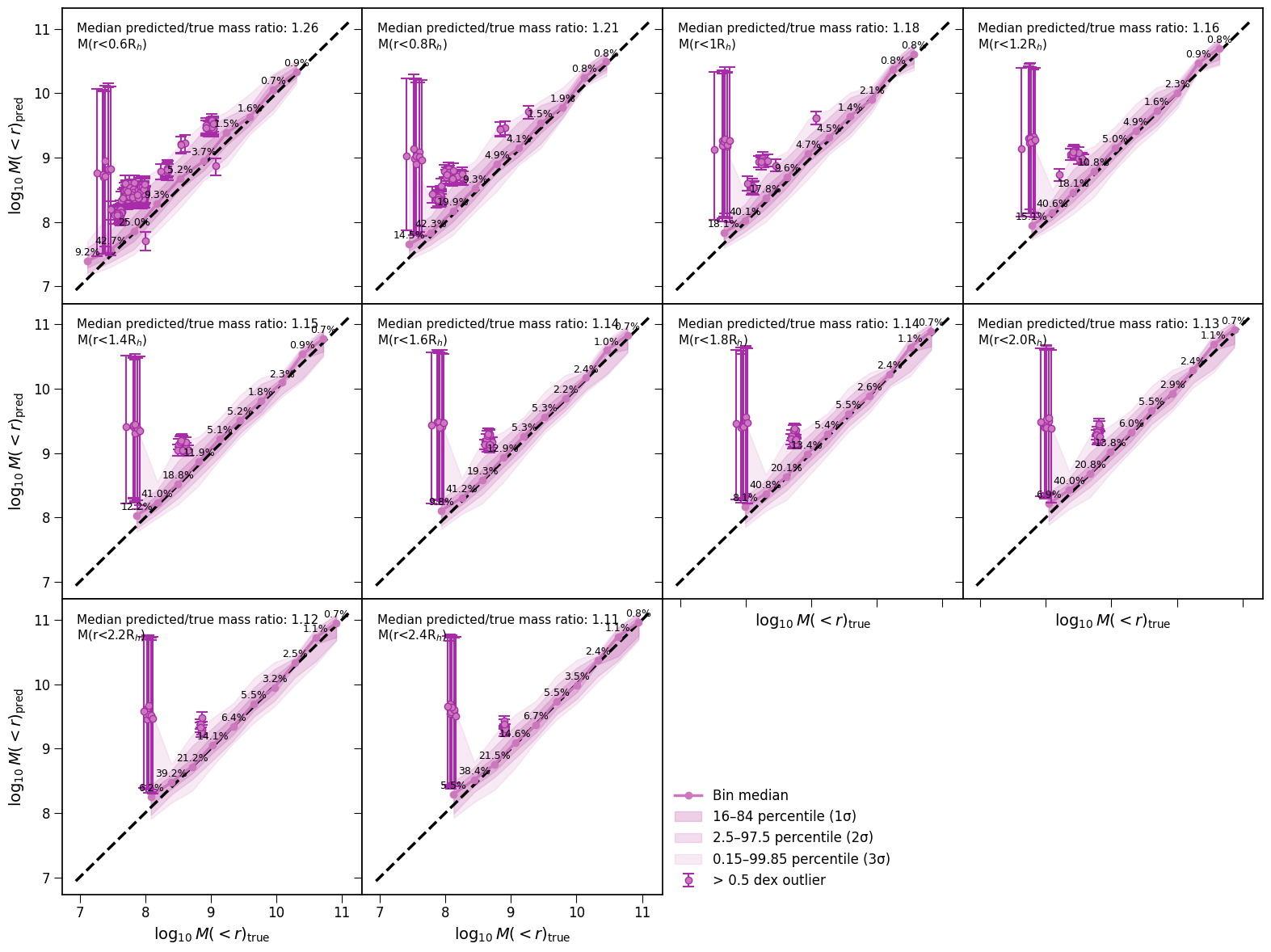}
    \caption{Comparison between predicted and true $\log_{10}$ enclosed mass for FIRE galaxies using the CNN+MAF model trained on the NIHAO+AURIGA dataset. Each panel corresponds to a different radius, indicated in the top-left corner. Circles connected by solid lines show the median predicted values across bins, with shaded regions denoting the 1, 2, and 3$\sigma$ scatter. The percentage of data within each bin is annotated above the corresponding circle. Outlier predictions, defined as deviations greater than 0.5 dex, are shown individually with error bars. The dashed black line marks the one-to-one relation. The median ratio of predicted to true enclosed mass (in linear space) is indicated in each panel. These values correspond to the dashed pink line in Fig. \ref{fig:ratio-test-FIRE-allmodels}.}
    \label{fig:FIRE_binned}
\end{figure*}

To better understand the deviations from the true values when predicting the FIRE testing set using our CNN+MAF model trained jointly on the NIHAO and AURIGA simulations, we plot the predicted $\log_{10}$ enclosed mass in each radial bin against the true values in Fig. \ref{fig:FIRE_binned}. This figure demonstrates that deviations are generally small in log space. However, we observe a tendency for the model to overpredict the masses of the least massive galaxies in the sample. This behaviour may reflect a bias introduced by the imbalance in our training dataset with respect to enclosed mass (see Fig. \ref{fig:mrdhist}). Since low-mass galaxies constitute a large fraction of the FIRE testing set, this bias significantly impacts the average model performance when evaluated on this dataset.

We also note a $\sim$1.5 dex overprediction for a subset of eight galaxy projections, all corresponding to the same galaxy (which appears 64 times in total, across different projections, in the testing set). Although we ensured that the individual input features (half-light radius, line-of-sight velocity dispersion, and the 98th percentile of line-of-sight velocities) fall within the limits of the training set, it remains possible that some galaxies lie outside the bounds of the training distribution when considering the joint input space. Indeed, we find that projections of this particular galaxy lie near the edges of the input space, suggesting the model is encountering input combinations it has not seen during training. This leads to significant prediction errors. Encouragingly, these outlier cases are associated with large uncertainties in the prediction, allowing us to identify when the inputs are too extreme for the model to produce reliable estimates.

As an alternative approach, we explored a stricter filtering procedure in which we retained only those testing projections that lie within the 99\% contours of the training data for every pairwise combination of input features. This method effectively removes the projections associated with large prediction errors, as well as other outliers. However, because these outliers do not significantly affect the overall model performance, we opted not to apply this cut. Instead, we retained all projections to preserve the instructive value of analysing cases where the model fails.

    \FloatBarrier
    \section{Additional posterior calibration checks}
    \label{app:additional-checks}

\begin{figure*}[ht]
        \includegraphics[width = \textwidth]{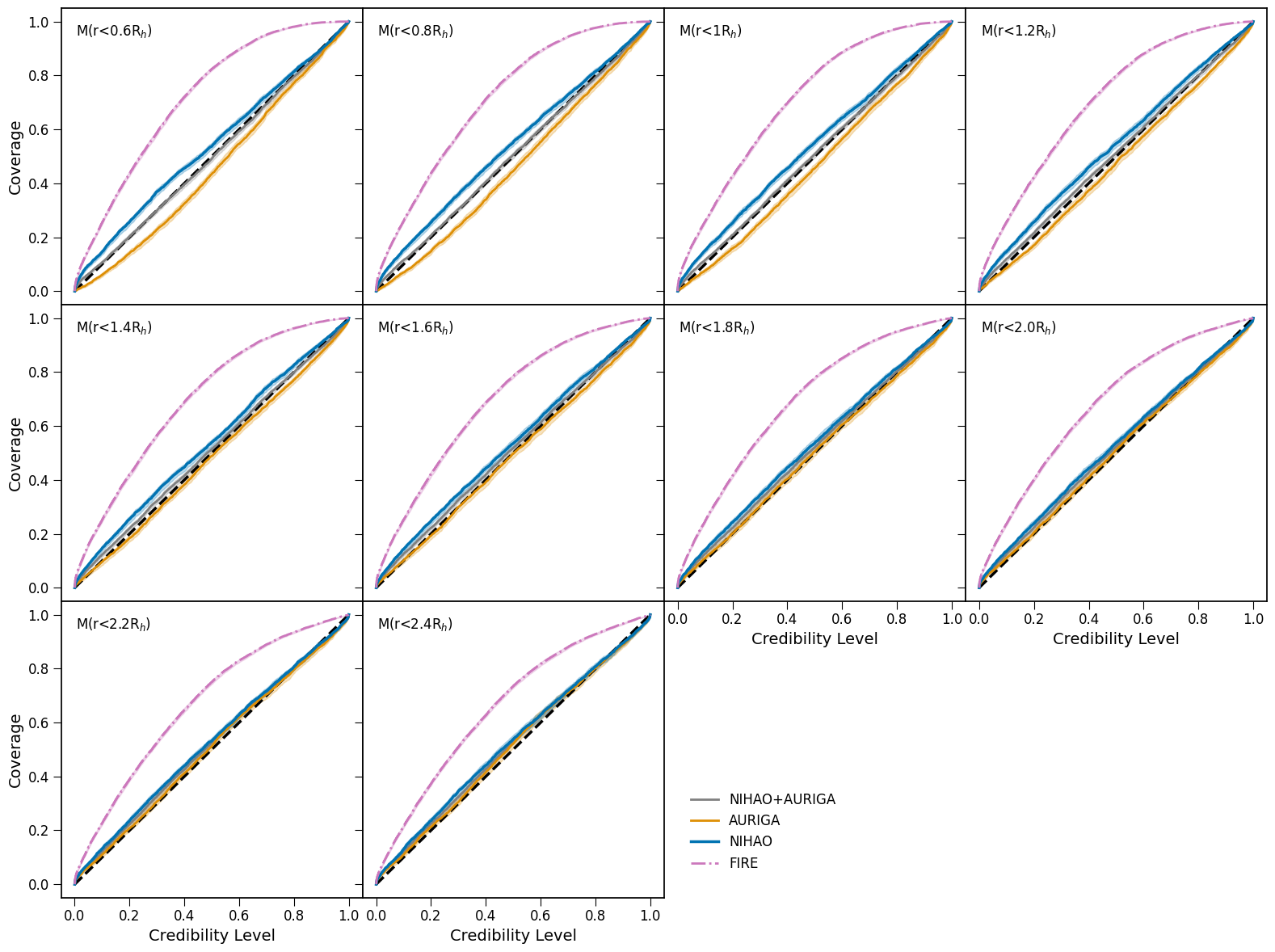}
  \caption{Marginal TARP test for all the dimensions of the posterior distributions generated by the CNN+MAF for the validation set of the model trained on NIHAO and AURIGA galaxies together. Results are also shown individually for galaxies of each suite in the validation set, together with results for the FIRE suite, which is not included in the training set.}
     \label{fig:perc_marginal_test}
    \end{figure*}

    We conducted marginal TARP tests on the model trained on NIHAO and AURIGA simulations simultaneously, assessing the calibration of the confidence intervals specifically for each dimension of the 10-d PDF. We show the tests for the validation set and for FIRE galaxy projections in \ref{fig:perc_marginal_test}. We find similar results to the TARP, demonstrating that the calibration of the posteriors is correct when marginalising over any of its dimensions when validating the model on simulations from the suites used to train it. The posterior distributions produced for FIRE galaxies show similar deviations in their calibration in all dimensions. 

    \begin{figure*}[ht]
    \centering
    \includegraphics[width=\columnwidth]{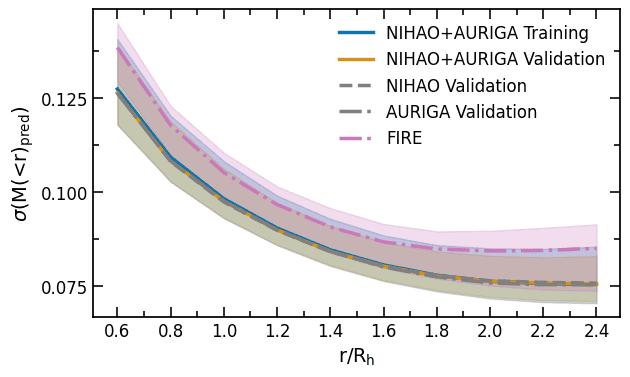}
    \caption{Uncertainty of the mass enclosed within different radii of the galaxies in the training set and in the validation set of the model trained on NIHAO and AURIGA galaxies together. The blue (orange) line shows the median uncertainty for the training (validation) set, while the shadowed region indicates the 1$\sigma$ interval. The median uncertainty of NIHAO and AURIGA galaxies in the validation set is individually shown as a dashed and dot-dashed grey line, respectively. The dotted pink line and shadowed region correspond to the FIRE testing set.}
    \label{fig:uncertainty-magnitude}
\end{figure*}

    After checking the calibration of the errors provided by our CNN+MAF model, we studied their magnitudes. In Fig. \ref{fig:uncertainty-magnitude} we plot the magnitude of the uncertainties estimated by our model for the log$_{10}$ value of the enclosed mass at different radii, for both the training and the validation set of the model trained on NIHAO and AURIGA simulations simultaneously, where we define the uncertainty as half the range between the 16th and 84th percentile of the enclosed mass at each radii according to the predicted posterior. The median uncertainty peaks for both the validation and training set at $\sim$0.125 for the mass enclosed within 0.6$\times$R$_{\rm h}$, and it decreases monotonically with radius, stabilising at $\sim$0.078 for masses within radii equal and larger than 1.8$\times$R$_{\rm h}$. We remind the reader that posterior distributions generated for FIRE galaxies are not correctly calibrated, and therefore the errors shown for this dataset are not representative of the actual deviation between the predicted and the truth value of the enclosed mass.

\end{appendix}

\label{LastPage}
\end{document}